\documentclass[reprint, superscriptaddress,
aps,
rmp,
]{revtex4-2}

\usepackage{amsmath,amssymb,amsfonts,latexsym}
\usepackage{dsfont}
\usepackage{bm} % bold symbols in math mode
\usepackage[mathcal]{euscript}
\usepackage{array}
\usepackage{multirow}
\usepackage{graphicx}
\usepackage{epsfig}
\usepackage{color}
\usepackage{tikz}
\usepackage[activate=normal]{pdfcprot}  % richtiges character protruding
\usepackage[colorlinks=true, linkcolor=blue, citecolor=blue]{hyperref}
\usepackage{siunitx}

\usepackage{psfrag}
\usepackage[caption = false]{subfig}

\newcommand{\gul}{Gulliver UMR CNRS 7083, ESPCI Paris, PSL Research University, 10 rue Vauquelin, 75005 Paris, France}
\newcommand{\puc}{Instituto de Física, Pontificia Universidad Católica de Chile, 8331150 Santiago, Chile}
\newcommand{\leiden}{Lorentz Institute, Leiden University, Niels Bohrweg 2, CA Leiden, 2333 The Netherlands}
\newcommand{\amolf}{AMOLF, Science Park 104, Amsterdam, 1098XG The Netherlands}
\newcommand{\bnu}{School of Systems Science, Beijing Normal University, Beijing, People’s Republic of China}
\newcommand{\nw}{Northwestern Institute on Complex Systems and ESAM,
Northwestern University, Evanston, IL 60208, United States of America}
\newcommand{\chh}{CHuepe Labs, 2713 West Augusta Blvd \#1, Chicago, IL 60622, United States of America}
\newcommand{\lyon}{Laboratoire de Physique, Univ Lyon, École Normale Supérieure de Lyon, CNRS, Lyon 69342, France}

\newcommand{\beq}{\begin{equation}}
\newcommand{\eeq}{\end{equation}}
\newcommand{\ben}{\begin{equation*}}
\newcommand{\een}{\end{equation*}}
\newcommand{\bseq}{\begin{subequations}}
\newcommand{\eseq}{\end{subequations}}
\newcommand{\bea}{\begin{eqnarray}}
\newcommand{\eea}{\end{eqnarray}}
\newcommand{\bal}{\begin{align}}
\newcommand{\eal}{\end{align}}

\newcommand{\ie}{\textit{i.e.} }

\newcommand{\mb}{\boldsymbol{m}}
\newcommand{\rb}{\boldsymbol{r}}
\newcommand{\ub}{\boldsymbol{u}}
\newcommand{\vb}{\boldsymbol{v}}
\newcommand{\nb}{\boldsymbol{n}}
\newcommand{\gb}{\boldsymbol{g}}
\newcommand{\fb}{\boldsymbol{f}}
\newcommand{\Fb}{\boldsymbol{F}}
\newcommand{\Fbext}{\boldsymbol{F}_{\rm ext}}

\newcommand{\Tsa}{\boldsymbol{T}_{\rm sa}}
\newcommand{\Tmt}{\boldsymbol{T}_{\rm ext}}
\newcommand{\wb}{\boldsymbol{w}}

\newcommand{\hxib}{\hat{\boldsymbol{\xi}}}
\newcommand{\hnb}{\hat{\boldsymbol{n}}}
\newcommand{\hvb}{\hat{\boldsymbol{v}}}
\newcommand{\heb}{\hat{\boldsymbol{e}}}

\newcommand{\apara}{\alpha_{\parallel}}

\newcommand{\Dr}{D_{\rm r}}

\begin{document}
\graphicspath{{./Figures/}}

\title{Self-Aligning Polar Active Matter}

\author{Paul Baconnier}
\affiliation{\gul}
\affiliation{\amolf}
\affiliation{\leiden}
\author{Olivier Dauchot} 
\affiliation{\gul}
\author{Vincent Démery}
\affiliation{\gul}
\affiliation{\lyon}
\author{Gustavo Düring}
\affiliation{\puc}
\author{Silke Henkes}
\affiliation{\leiden}
\author{Cristián Huepe}
\affiliation{\bnu}
\affiliation{\nw}
\affiliation{\chh}
\author{Amir Shee}
\affiliation{\nw}

\date{\today{}}

% MAIN TEXT ===================================================================================================

\begin{abstract}
Self-alignment describes the property of a polar active unit to align or anti-align its orientation towards its velocity. In contrast to mutual alignment, where the headings of multiple active units tend to directly align to each other -- as in the celebrated Vicsek model --, self-alignment impacts the dynamics at the individual level by coupling the rotation and displacements of each active unit. This enriches the dynamics even without interactions or external forces, and allows, for example, a single self-propelled particle to orbit in a harmonic potential. 
At the collective level, self-alignment modifies the nature of the transition to collective motion already in the mean field description, and it can also lead to other forms of self-organization such as collective actuation in dense or solid elastic assemblies of active units. This has significant implications for the study of dense biological systems, metamaterials, and swarm robotics.
Here, we review a number of models that were introduced independently to describe the previously overlooked property of self-alignment and identify some of its experimental realizations. Our aim is three-fold: (i)~underline the importance of self-alignment in active systems, especially in the context of dense populations of active units and active solids; (ii)~provide a unified mathematical and conceptual framework for the description of self-aligning systems; (iii)~discuss the common features and specific differences of the existing models of self-alignment. We conclude by discussing promising research avenues in which the concept of self-alignment could play a significant role.
\end{abstract}

%\pacs{}

\maketitle

\tableofcontents{}

\section{Introduction}
\label{intro}
Polar active matter encompasses a class of systems composed of self-driven units that convert energy into directed motion or forces. The microscopic degrees of freedom are two-fold: those describing the position of the active units, and those describing the orientations or polarities of their self-propulsion forces.

When the interactions between active units do not affect their orientations, as they only couple the positional degrees of freedom, and are typically given by the usual attraction and/or repulsion forces, one talks about \emph{scalar} active matter~\cite{Wittkowski2014}.
Even so, scalar active matter can exhibit interesting distinct collective phenomena, such as the emergence of Motility Induced Phase Separation (MIPS)~\cite{Cates2015} in systems of active Brownian particles.

In more general cases, the presence of orientational degrees of freedom due to the polarity of active units allows for additional types of interactions. Two classes of such interactions can be distinguished, depending on their symmetries: \emph{polar} and \emph{nematic} mutual alignment.
Polar alignment is said to take place when the polarities of two particles interact like the spins of an XY model, mostly through some kind of ferromagnetic-like coupling~\cite{Zhao2021}. This is the type of alignment that was introduced in the seminal Vicsek model~\cite{Vicsek1995}. 
Instead, nematic alignment is said to take place when the interaction is agnostic to the sign of the polarity, as in liquid crystals. 
Polar alignment is known to drive collective motion~\cite{Vicsek2012}, while nematic alignment has been linked to the so-called ``active turbulence''~\cite{Alert2022}.
Note that both forms of mutual alignment require an explicit coupling between the orientations of the interacting units.
General reviews on active systems with mutual alignment can be found in~\cite{Toner2005, Juelicher2007, Ramaswamy2010, Marchetti2013}.

A different and mostly overlooked way to introduce interactions that involve the polarity of active units is to consider the possibility of couplings between orientations and translations.
One particular case of interest is \emph{self-alignment}, where the orientation of each polar particle couples to its own translational degrees of freedom, more specifically to its velocity.
Self-alignment was introduced as early as 1996, in pioneering work by~\cite{Shimoyama1996}, stemming from the basic observation that the polar self-driven headings and velocities of the active units do not have to be parallel. It was only ten years later that it was reintroduced in~\cite{Szabo2006} to describe the collective migration of tissue cells. 

The next decade saw the sporadic reintroduction of self-alignment in a variety of contexts. 
It was used to study active jamming  in~\cite{Henkes2011}, motivated by in vitro experiments on confluent monolayers of migratory epithelial and endothelial cells. 
It was part of a decentralized control algorithm for a wheeled robot swarm formation implemented in ~\cite{Ferrante2012} that led to a minimal model of active elastic crystals and solids composed of self-driven agents linked by permanent springs~\cite{Ferrante2013}. 
It was found to result from the mechanical dynamics of an experimental system of self-propelled vibrated polar disks and shown to be a key ingredient for the emergence of its collective motion~\cite{Deseigne2010, Weber2013}.

It is only recently that self-alignment has started to attract more attention, especially in the context of dense and solid active matter, where it often leads to collective dynamics that are distinct from those observed for polar mutual alignment. Indeed, although it was long assumed that self-alignment was a curiosity of the single-particle dynamics that could be treated as effective mutual alignment at the collective level, it has now been shown that novel collective effects can emerge if the structure of the assembly remains frozen at long timescales due to confinement or cohesion, when the system behaves like an elastic solid rather than like a viscous liquid.

In the context of epithelial cell sheets, for example, it has been shown that self-alignment can induce collective oscillations in Voronoi vertex models~\cite{Barton2017, Petrolli2019} and phase field models~\cite{Peyret2019}. In dense biofilms where the active units can be modeled as embedded in an elastic network, it can also lead to various forms of collective oscillations~\cite{Xu2023}.
On the other hand, in artificially designed active elastic metamaterials where self-driven components are coupled through a network of mechanical links, self-alignment can lead to different forms of collective motion and collective actuation (in which agents perform large scale oscillations around a reference configuration) that strongly depend on the network properties~ \cite{Ferrante2013a,Woodhouse2018, Baconnier2022, Zheng2023, Turgut2020}.
In addition, recent work has shown a complex interplay between self-alignment and the active glass transition~\cite{Paoluzzi2022}.
Finally, self-alignment has increasing applications for decentralized control in the context of swarm robotics, due to its emergent collective dynamics and potentially simple implementation \cite{Zheng2020,BenZion2023}.

The purpose of this review is to provide a unifying overview on the various phenomena, underlying mechanisms, emergent features, and applications related to self-alignment in different fields of active matter.
More specifically, our first goal is to underline the importance of self-alignment, especially in the context of dense and solid active units. We describe the potential mechanisms that can lead to self-alignment and discuss current experiments that demonstrate the role of self-alignment in emerging collective phenomena. Our second goal is to provide a common mathematical framework for the description of self-aligning systems. Finally reviewing the existing models that include self-alignment, we aim at discussing their similarities and differences, in the context of this common framework.

The review is organized as follows. 
In Section~\ref{sec:context}, we will formally define the concept of self-alignment and present the general forms of its equations of motion for active agents.
Section~\ref{sec:singlepart} describes the experimental and theoretical works that have analyzed the self-aligning dynamics of single agents, and discusses different ways in which self-alignment can be implemented.
In Section~\ref{sec:collmot} we review different situations where self-alignment leads to collective motion in active liquids, in the absence of other sources of mutual alignment between the agents. 
Section~\ref{sec:collact} describes various experimental and numerical contexts in which collective actuation has been observed in dense or solid active systems with self-aligning dynamics.
Finally in Section~\ref{sec:discussion}, we discuss unifying perspectives that connect the different types of self-aligning dynamics presented in this review with each other and with other novel forms of active dynamics that have been introduced in the literature. 

% ============================================
\section{Self-alignment: a torque coupling orientation and displacement}
\label{sec:context}
% ===============================
\subsection{Symmetry considerations}

\begin{figure}[b!]
\vspace{-0mm}
\center
\includegraphics[width=0.67\columnwidth,trim = 0mm 0mm 0mm 0mm, clip]{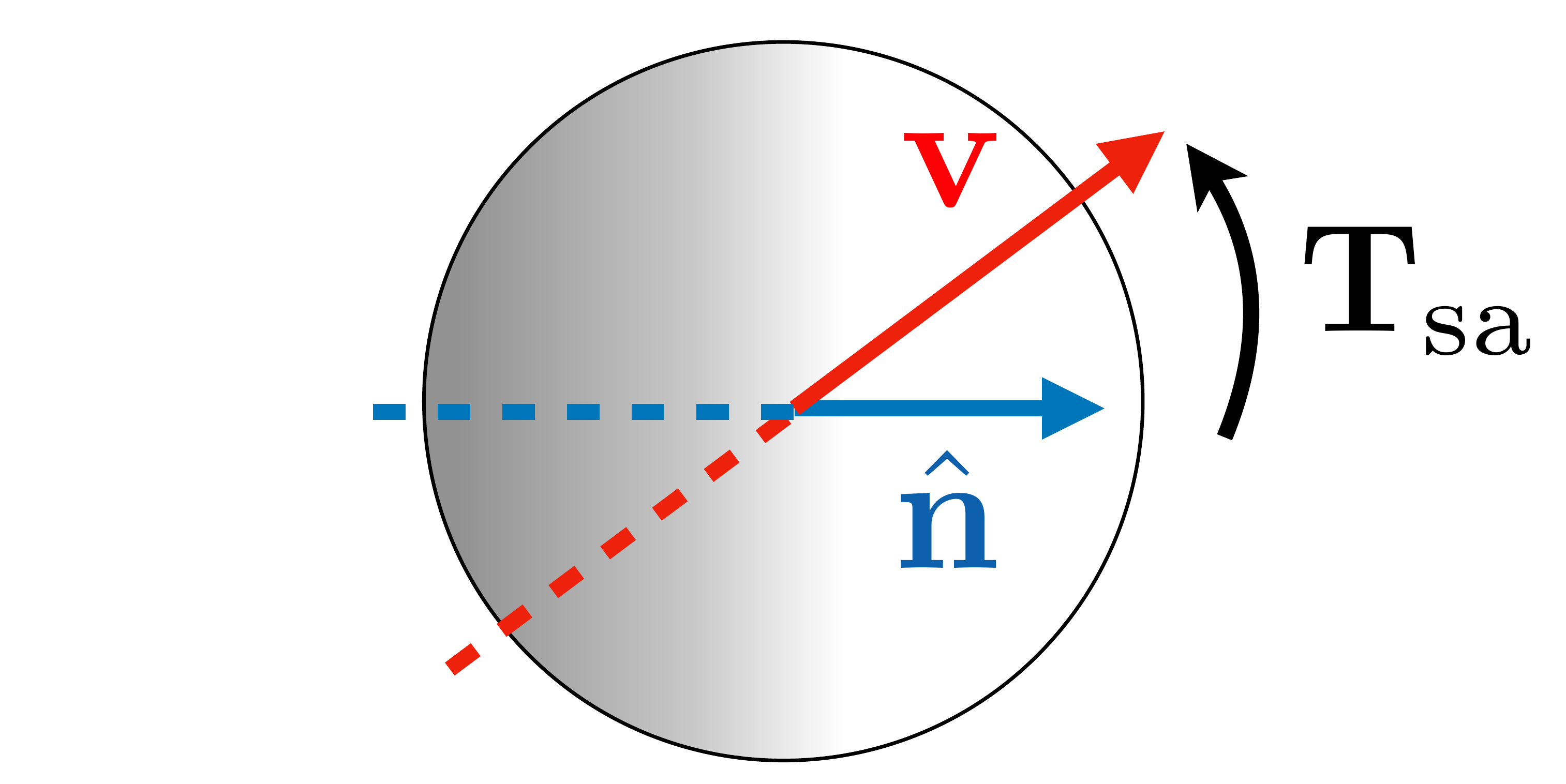}
\vspace{-0mm}
\caption{{\bf Schematic representation of self-alignment:} The grey shaded zone, represents the distribution of the propulsive and dissipative forces acting on the isotropic body. It defines the polar axis $\hnb$ and is left-right symmetric with respect to it by design. When the velocity $\vb = \dot\rb$ is not aligned with $\hnb$, a self-aligning torque results from the asymmetric distribution of the propulsive and dissipative forces with respect to $\vb$.}
\label{fig:SA-sketch}
\vspace{-5mm}
\end{figure}

A precise mechanical argument for the origin of self-alignment will be provided in Section~\ref{sec:discussion} using a mechanical walker model. Let us here provide a first intuition of it, following simple symmetry considerations. Suppose one aims at designing the simplest possible self-propelled polar agent. While the shape of its body could remain isotropic (circular in 2d, spherical in 3d), one would need to embed a unit vector $\hnb$ describing the tail-head axis of the agent, also called polar axis, along which it gains momentum due to self-propulsion (Fig.~\ref{fig:SA-sketch}).
In order to ensure an unbiased motion of the agent, the design needs to respect the axial symmetry of the body. 
This includes the distribution of mass, but also the spatial distribution of propulsive and dissipative forces generated at its interface with the surroundings, so that no systematic torque is exerted on the agent when it moves along its polar axis. Note that any internal active or passive forces are subject to action-reaction and don't contribute if the agent retains its shape.
It is then clear that the very same distribution of propulsive and dissipative forces are not axially symmetric with respect to the direction of motion, as soon as it is misaligned with the polar axis.
This asymmetric distribution of forces generically exerts a torque on the agent body, which thus rotates its polar axis towards the direction of motion or its opposite; this is self-alignment.

In the case of the vibrated polar disks introduced in~\cite{Deseigne2010}, these have two contacts with the vibrating ground: a large piece of rubber and a thin metallic tip (Fig.~\ref{fig:collmot_phenomenology}). The rubber piece is responsible for most of the friction. As soon as the disk moves in a direction that is not aligned with its polar axis, the asymmetric friction on the ground makes it rotate towards its direction of motion. 
In the case of the Hexbugs~\copyright (Fig.~\ref{fig:single_particle}-$b_1$), or the bristle-bots~ \cite{Giomi2013}, legs are bent towards the front. While it is more difficult to separate the dissipative from the propulsive elements, the above generic symmetry argument still holds. 

Interestingly the sign of the self-aligning torque depends on the distribution of the friction forces: when the agent experiences more friction on the back than on the front, it spontaneously aligns toward its velocity, while it anti-aligns in the opposite case. It is therefore possible to set the sign of the self-alignment by design~\cite{BenZion2023}.

\begin{table*}[t!]
%\hspace*{-0.3cm}
\begin{tabular}{p{0.32\textwidth}|>{\centering}p{0.1\textwidth}|>{\centering}p{0.1\textwidth}|>{\centering}p{0.1\textwidth}|>{\centering}p{0.1\textwidth}|>{\arraybackslash}p{0.2\textwidth}} 
Reference &  $ \wb $ & L / NL  & Isotropic Damping & Confinement & Type of System
\tabularnewline
\hline
\cite{Shimoyama1996}   &  $\hvb$  &  NL  & Y & N & Particles
\tabularnewline
\cite{Szabo2006}       &  $\hvb$  &  L & Y & N &Particles
\tabularnewline
\cite{Henkes2011}      & $\hvb$   &  L   & Y & Y & Particles
\tabularnewline
\cite{Barton2017} & $\hvb$ & NL  & Y & Y & Voronoi
\tabularnewline
\cite{Malinverno2017}  & $\hvb$   & L & Y & N & Voronoi
\tabularnewline
\cite{Giavazzi2018}    &  $\hvb$  & NL & Y & N & Voronoi
\tabularnewline
\cite{Petrolli2019}    & $\hvb$    & NL & Y & Y & Voronoi
\tabularnewline
\hline
\cite{Weber2013}     & $\vb$    &  NL  & Y & N & Particles
\tabularnewline
\cite{Ferrante2013}  & $\vb$    &  NL  & N & N & Robots or structure
\tabularnewline
\cite{Lam2015}       & $\vb$    &  NL  & Y & N & Particles
\tabularnewline
\cite{Dauchot2019}    &  $\vb$   & NL & Y & Y & Single particle
\tabularnewline
\cite{Peyret2019}     &  $\vb$   & L & Y & Y & Cells or Phase-field
\tabularnewline
\cite{Lin2021}        & $\vb$    &  NL  & N & N & Robots or structure 
\tabularnewline
\cite{Baconnier2022} & $\vb$     & NL & Y & Y & Structure
\tabularnewline
\cite{Damascena2022} & $\vb$     & NL & Y & Y & Single particle
\tabularnewline
\cite{Xu2023}        & $\vb$     & NL & N & Y & Bacteria
\tabularnewline
\hline 
\end{tabular}
\caption[table caption]{\small{\textbf{Self-alignment interactions as defined in the literature: the papers are grouped in two sets according to the choice of $\wb = \hvb$ or $\wb = \vb$}, and sorted chronologically in each set. The $3^{rd}, 4^{th}$ and $5^{th}$ columns indicate whether the self-aligning coupling is linear (L) or non linear (NL)}, whether the damping is isotropic (Y) or anisotropic (N), whether the system is studied under confinement (Y) or no (N). Finally the last column specifies the type of ``agent/system" considered.}
\label{table:all_models}
\end{table*}

\subsection{Equations of motion} 
In two dimensions, the deterministic motion of a self-aligning polar agent with self-propelled heading along $\hnb$ can be described by Newton's equations %
\bseq \label{eq:New}
\begin{align} 
	m \ddot\rb &= F_a \hnb - \gamma \dot\rb + \Fb_{\rm ext}(\rb), \label{eq:New_trans}\\
	J   \ddot{\hnb} 	&=  \Tsa \times \hnb - \gamma_r \dot{\hnb},  \label{eq:New_rot}
\end{align}
\eseq
where $m$ and $J$ are respectively the mass and inertial momentum of the agent, while $\gamma$ and $\gamma_r$ encode the respective translational and rotational damping. These are usually taken as scalar, but can also be tensorial for non-symmetric shapes. Equation (\ref{eq:New_trans}) describes the inertial translational motion of an agent that is self-propelled by an active force $F_a \hnb$ and is subjected to an external force $\Fb_{\rm ext}(\rb)$. Equation (\ref{eq:New_rot}) expresses self-alignment in the form of a torque $\Tsa$ that couples the orientation $\hnb$ of the self-propulsive force to the velocity $\dot\rb$ of the agent. We stress again that this self-alignment is distinct from a mutual Vicsek alignment.
Note also that it is assumed here that there is no additional torque acting on the body, so that we restrict our discussion to a-chiral active matter.

As we shall discuss further below, the torque can be chosen to be proportional to the magnitude of the velocity or to solely depend on its orientation, as expressed by defining either $\wb = \vb=\dot\rb$ or $\wb = \hvb = \dot\rb/|\dot\rb|$ in the following expression:
\beq
\label{eq:Tsa}
\Tsa = \zeta (\hnb \times \wb),
\eeq
where $\zeta$ is the coupling coefficient. 

The inertial form of Eqs.~(\ref{eq:New}) has been rarely used for describing self-aligning systems. Indeed, other than in~\cite{fersula2024self}, where the authors report unexpected effects of the coupling between inertia and self alignment,  the angular dynamics have always been assumed to be overdamped,  $J/\gamma_r\rightarrow 0$. The reason is that, for all agents considered so far, the rotational damping has been rather large because of the shape factor of the agents. 
In what follows, we will therefore restrict ourselves to this limit and use the equations of motion
\bseq \label{eq:New_aod}
\begin{align} 
	m \ddot\rb &= F_a \hnb - \gamma \dot\rb + \Fb_{\rm ext}(\rb), \label{eq:New_trans_ud}\\
	\dot{\hnb}	 &= \beta (\hnb \times \wb) \times \hnb.  \label{eq:New_rot_od}
\end{align}
\eseq
In the case with $\wb = \vb$, where the coupling is proportional to the magnitude of the velocity, $\beta = \zeta/\gamma_r$ has the dimension of an inverse length-scale $l_a^{-1}$, which we call the alignment length. In the case $\wb = \hvb$, where the coupling is independent of the magnitude of the velocity, $\beta$, has the dimension of an inverse timescale $\tau_a^{-1}$, which we call the alignment time.

Another commonly used variant in the angular dynamics consists in replacing it by a simple linear relaxation towards the velocity orientation $\psi$, as in~\cite{Szabo2006, Henkes2011}. Here, $\dot\theta=\beta(\psi-\theta)$, where $\theta$ defines the orientation of the polar vector $\hnb$. Note, however, that this is not equivalent to a linearization of the full system of equations~\ref{eq:New_aod}.
%, for a small angle difference $\psi-\theta$.

Finally, in the most commonly used form of the equations of motion, the translational dynamics is also considered to be in the overdamped limit and noise terms are added to the deterministic equations. These noise terms need not satisfy Einstein's relation since the agent is intrinsically out of equilibrium and the noise can thus take its source from the driving or from the dissipative interactions, in addition to the usual thermal bath, when significant. The translational noise does not need to be isotropic either. 
In practice, in most cases the translational noise is neglected and all noise distributions are chosen to be Gaussian, delta-correlated, and isotropic. If we further consider the translational damping to also be isotropic, we can replace the tensor $\gamma$ by a scalar coefficient $\gamma I$, and obtain the equations:
\bseq \label{eq:New_od}
\begin{align} 
	\dot\rb &= v_0 \hnb + \frac{1}{\gamma}\Fb_{\rm ext}(\rb), \label{eq:New_trans_od}\\
	\dot{\hnb}	 &= \beta (\hnb \times \wb) \times \hnb + \sqrt{2D_r}\eta \hnb^{\perp},  \label{eq:New_rot_od_noise}
\end{align}
\eseq
where $v_0 = F_a/\gamma$ and $\eta$ is a random variable that introduces Gaussian white noise with unit variance. In the absence of self-alignment ($\beta = 0$), one therefore recovers the standard Active Brownian Particle  model.

Table~\ref{table:all_models} provides a synthetic view of the main articles where self-alignment was introduced in the literature that categorizes the variations in the equations of motion considered in each case.
We group them according to whether $\wb=\vb$ or $\hvb$, whether the angular dynamics is linear, and whether the translational damping is isotropic.

\section{Individual self-aligning agent dynamics}
\label{sec:singlepart}

\subsection{Simple experiments with mechanical walkers} 
\label{subsec:walkers} 

\begin{figure*}
\includegraphics[width=\textwidth]{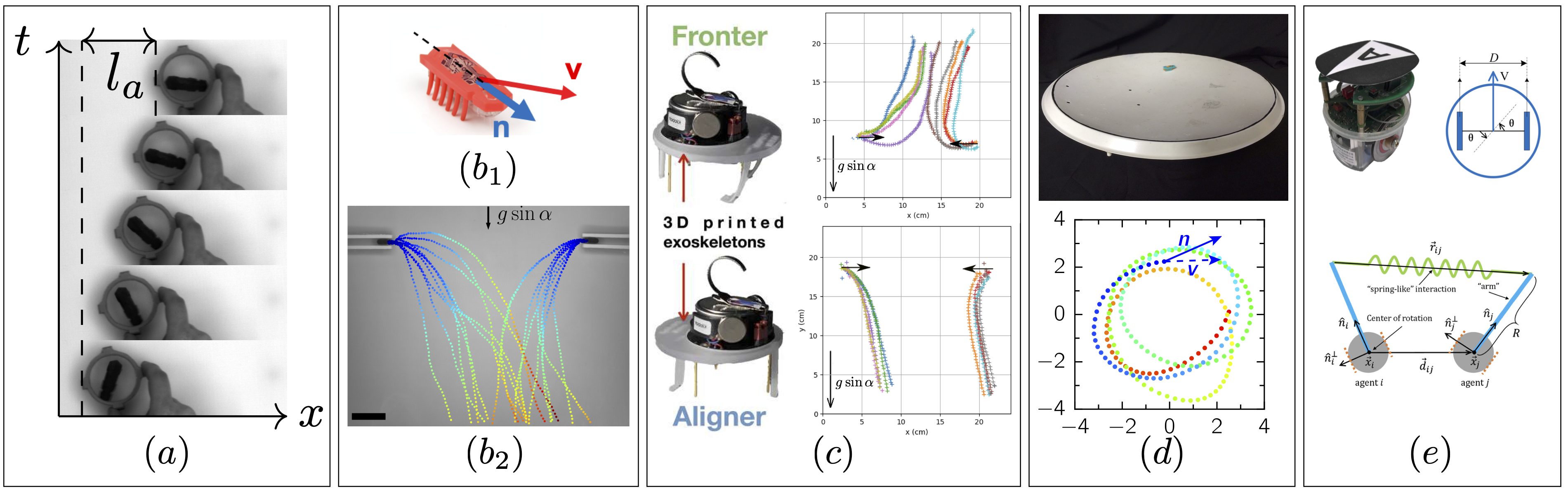}
\caption{
\textbf{Self-alignment in dynamics of single walkers} (Color online) (a) When a self-aligning agent, here a Hexbug\copyright (see b-Top), is manually translated in a direction different from the one of its tail-head polarity, while leaving free to reorient, its orientation relaxes toward the direction of the imposed motion, on a characteristic distance $l_a$, adapted from~\cite{Baconnier2023}; 
(b) Top: a hexbug, with its velocity $\hvb$ that is not necessarily align with its tail-head polarity $\hnb$. Bottom: when moving on an plane inclined of an angle $\alpha$ with respect to the horizontal, a hexbug aligns in the direction of the gravity force, adapted from~\cite{Baconnier2023}; 
(c) A kilobot~\cite{Rubenstein2014} embedded into two morphologically distinct 3d printed exoskeleton align either towards (Bottom) or against (Top) the direction of the gravity force, denoting a positive, respectively a negative value of $\beta$ (adapted from~\cite{BenZion2023} with additional data); 
(d) Orbiting dynamics of a hexbug in a parabolic dish (adapted from~\cite{Dauchot2019}); 
(e) Bottom: schematic representation of the active elastic model for two interacting self-propelled agents. Each agent is represented by a gray disk with a blue ‘arm’ projecting forward a distance $R$. The green sinusoidal line represents a linear spring connecting the tips of these arms. The agent positions and centers of rotations are given by $x_i$ and  $x_j$ while their heading polarities are indicated by the unit vectors $\hnb_i$ and $\hnb_j$, respectively, (adapted from~\cite{Lin2021}). The orange dotted lines show the position of wheels in a potential mechanical realization with actual robots such as the one shown on Top~\cite{Zheng2020}.
}
\label{fig:single_particle}
\end{figure*}

Vibrated or vibrating mechanical walkers, which are often used in the experimental study of polar active matter~\cite{Deseigne2010,Deseigne2012,Giomi2013,Baconnier2022,BenZion2023}, generically exhibit self-alignment. This can be demonstrated in a simple experiment that imposes the translational velocity of the walker while leaving its polarity free to reorient. If the walker exhibits self-alignment, it will reorient along the imposed direction of motion. Such an experiment was conducted in~\cite{Baconnier2022} in order to extract the parameters of the dynamics of individual Hexbugs (see Fig.~\ref{fig:single_particle}-a). Using a simple mechanical device to drive the Hexbug along a square trajectory while allowing it to rotate freely, the authors demonstrated that, for this specific system, $\wb = \dot\rb$ and self-alignment is controlled by an alignment length. Indeed, the angle between the polarity vector and the velocity vector decreases exponentially to zero after each change of imposed translational direction  with a characteristic time $\tau_a(V)\sim 1/V$, where $V$ is the imposed translational speed. This results in the measurement of an alignment length $l_a = V \tau_a(V)$ that in this case is found to be of the order of one Hexbug body-length.

An alternative manifestation of self-alignment can be observed when a constant external force is applied to the polar walker. In that case, the velocity relaxes towards the sum of the active force and the external one. Since the polarity tends to align with the velocity, it will converge towards aligning with the external force. 
Experimentally, the simplest implementation of an external force consists in placing the active agent on a plane inclined by a small angle $\alpha$ with respect to the horizontal, thereby imposing a constant force $g\sin\alpha$ downhill. A Hexbug placed on such a tilted plane will therefore reorient its body downwards along the slope (Fig.~\ref{fig:single_particle}-b).
Note that the sign of the self-aligning torque is determined by the design of the walker's morphology. It can be positive, as it is the case for Hexbugs, or negative, as recently illustrated in~\cite{BenZion2023} for the case of augmented Kilobots (Fig.~\ref{fig:single_particle}-c). 

\subsection{A self-aligning walker in a harmonic potential}
\label{subsec:harmonicpot}
The importance of self-alignment is best illustrated when considering the dynamics of a self-aligning active particle in a harmonic potential. This canonical problem was studied in~\cite{Dauchot2019} while exploring the dynamics of a Hexbug in a parabolic dish antenna (Fig.~\ref{fig:single_particle}-d). The deterministic dynamics of this system is described by Eq.~(\ref{eq:New_aod}) in the $\wb=\vb$ case with external force $\Fb_{\rm ext} = -k \rb$, where $k$ is the stiffness of the potential and $\rb$ is the particle position with respect to its minimum. Note that here, as in Eq.~(\ref{eq:New_aod}), the translational damping is assumed to be isotropic, since the effect of the external forces on the velocity is independent of particle orientation. A more detailed derivation of the dynamics in both the isotropic and anisotropic cases is included in Appendix \ref{app:harmonic_potential} and \ref{app:anisotropic_damping}, respectively.

We can rescale the dynamical equations by a characteristic length scale $l_e = F_a/k$, the ``elastic length'' at which the harmonic potential force is balanced by the active force, and by a characteristic timescale $t_0 = \gamma/k$. The resulting dimensionless equations of motion in the noiseless case then become
\bseq \label{eq:Demery}
\begin{align}
	\tau_v \ddot\rb &= \hnb - \dot\rb  - \rb, \\
	\tau_n \dot{\hnb} &= (\hnb \times \dot\rb) \times \hnb, \label{eq:sa_rot}
\end{align}
\eseq
with $\tau_v = mk/\gamma^2$ and $\tau_n = k/(\beta F_a)=l_a/l_e$. 
In the overdamped ($\tau_v = 0$) case, these equations admit an infinite set of fixed points ($\dot\rb=0$, $\dot\nb=0$), where the agent sits at distance $r = 1$ from the origin, pointing radially, with $\rb = \hat\nb$. In the presence of noise, the agent diffuses along this continuous set of fixed points (see Fig.~\ref{fig:single_particle}). These states, referred to as ``climbing'' states in~\cite{Dauchot2019}, are marginally stable for $\tau_n\geq 1$ and unstable for $\tau_n<1$.
In this latter case, one finds instead a stable ``orbiting'' state, in which the active agent rotates in the trap along a circular orbit of radius $r=\sqrt{\tau_n}$ and tangential velocity $v_\theta = r \omega =\sqrt{1-\tau_n}$, where  $\omega$ is the angular velocity.
The rotation is clockwise or counterclockwise, spontaneously breaking the chiral symmetry of the equation of motions.
At the transition from the climbing state to the orbiting state, the amplitude of the oscillations is finite and the frequency is zero. This corresponds to a drift-pitchfork bifurcation~\cite{Kness1992}, which was initially analyzed in the context of reaction-diffusion models.

Remarkably, inertia does not affect the stability of the climbing state, which remains marginally stable for $\tau_n\geq 1$. Inertia also does not affect the nature of the transition as long as $\tau_v\leq 1$. If we increase inertia beyond this threshold, the stable orbiting state appears at $\tau_n$ values below $\tau_n^*=\tau_v/[2\sqrt{\tau_v}-1]\geq 1$. Hence, the climbing and orbiting states coexist for $1\leq\tau_n\leq\tau_n^*$ and the transition becomes discontinuous.

In the overdamped limit, the above picture has been extended to the case of an anisotropic harmonic trap~\cite{Damascena2022}.
In this case the fixed points still form a loop, topologically speaking, but not a unit circle. They do not become unstable at the same level of activity: the fixed points in the stiffest parts of the trap destabilize first, and those in the shallowest parts destabilize last.
The dynamical regimes are also more complex: instead of circles, the particle follows an oval for a weakly asymmetric potential, which deforms to a lemniscate and then to higher orders lemniscates upon increasing the asymmetry. There are two interesting limiting cases. When the soft direction becomes flat, the active unit polarizes in this direction and starts unbounded motion. When the hard direction becomes infinitely rigid, the orbiting state does not exist. Interestingly, it was also shown that in this last case the angular noise can still restore pseudo periodic dynamics~\cite{Baconnier2023}. 

%Reproducing the above analysis for $\wb = \hvb$, one notes that there are important qualitative differences, already in the overdamped limit. The characteristic time scale of the angular dynamics becomes agnostic to the intensity of the active force : it is simply the dimensionless alignment time $\hat{\tau}_a =\hat{\beta}^{-1} = k \tau_a / \gamma$. The climbing state is always unstable and the orbiting state exists and is stable for all $\hat{\beta} > 0$. The orbit takes place at distance $r$ from the origin, with a tangential velocity $v_\theta = r \omega$, where:    
%\bseq
%\begin{align}
%r & = \hat{\tau}_a \sqrt{\frac{\sqrt{1+4\hat{\tau}_a^{-2}}-1}{2}},\\
%v_\theta & = \hat{\tau}_a \frac{\sqrt{1+4\hat{\tau}_a^{-2}}-1}{2},
%\end{align}
%\eseq
%

Finally, we note that the case of anti-aligning active particles in a harmonic potential was reported by~\cite{BenZion2023}. In this case, the climbing state becomes unconditionally stable and no orbiting solution emerges.
%\vd{Same as ABP. In the absence of external forces, it is the same as ABP for $m=0$. At first order in $m$, the mass renormalizes the diffusion coefficient. I do not know what happens at higher orders. Should we talk about that?}

\subsection{Other realizations of self-alignment}

\subsubsection{Off-centered mechanical forces}
\label{subsubsectionOffcentered}
A specific type of mechanical interaction that can lead to dynamics equivalent to self-alignment was considered in ~\cite{Lin2021}. In this case, each active agent is subject to off-centered forces applied on a point $\rb + R \hnb$, at the end of a lever arm of length $R$ that projects forward from the center of rotation $\rb$ (see Fig.~\ref{fig:single_particle}-e).
These forces will introduce a mechanical torque equal to
\beq
%\Tmt = R [\hnb \times \Fb_{\rm ext}(\rb + R \hnb)] \times \hnb.
\Tmt = R \hnb \times \Fb_{\rm ext}(\rb + R \hnb).
\label{eq:Text}
\eeq
The corresponding Newton's equations for these agents are thus equivalent to those in Eq.~(\ref{eq:New}), but replacing the self-aligning torque $\Tsa$ by $\Tmt$.
Note, however, that now the forces are not computed as only a function of particle positions; they will also depend on their orientations since the argument of $\Fb_{\rm ext}$ includes $\hnb$. This is because the interaction forces now depend on the distance from lever arm to lever arm, rather than from center of rotation to center of rotation.

It was shown in \cite{Lin2021} and \cite{Lin2023} that this dependency of $\Fb_{\rm ext}$ on $R \hnb$ results in several effects that are not found in self-aligning dynamics. For example, it allows the interaction to be set as mainly based on mutual alignment, for large $R$, or on self-alignment, for small $R$. It also results in states of quenched disorder that cannot form in self-aligning systems.

In a case with overdamped dynamics and isotropic translational damping, Eq.~(\ref{eq:New_trans_od}) becomes 
$\dot\rb = v_0 \hnb + (1/\gamma) \Fb_{\rm ext}(\rb + R \hnb)$. This can be inserted into Eq.~(\ref{eq:Text}) to obtain an expression for the torque akin to the self-alignment torque $\Tsa$ in Eq.~(\ref{eq:Tsa}) for the $\wb = \dot\rb$ case, if we define $\zeta = \gamma R$.
In the limit of small $R$, the resulting dynamics will then become identical to self-alignment.
%it can be used to substitute $\dot\rb$ in the expression for $\Tsa$, so that the off-centered agent dynamics and the self-aligning one in the case $\wb = \dot\rb$ are now equivalent with $\beta = R \tgam / \gamma_r $.
On the other hand, if the translational damping is anisotropic, this equivalence breaks down since the expression corresponding to Eq.~(\ref{eq:New_trans_od}) will be $\gamma \dot\rb = F_a \hnb + \Fb_{\rm ext}(\rb + R \hnb)$, where $\gamma$ remains a tensor. 
In the limiting case with infinite translational damping along $\hnb^{\perp}$, where no lateral displacements are allowed (as in the models of wheeled agents reviewed in section~\ref{ABES}), this difference is crucial because $\hnb \times \wb = 0$ and the self-alignment torque in Eq.~(\ref{eq:New_rot_od}) would vanish. By contrast, even in this limit, models based on off-centered forces will produce angular dynamics that resemble self-alignment, since $\Tmt$ will not vanish.

\subsubsection{Microswimmers}
\label{subsubsectionJanus}
Another potential realization of self-alignment dynamics can be found in Microswimmers, typically defined as micron-sized self-propelled entities in suspension in a solvent. These take their momentum from the surrounding fluid while conserving the total momentum. In such a situation, one may wonder whether the coupling of the translational and rotational degrees of freedom, which gives rise to self-alignment, will subsist. 
This question was already raised for the simplest case of asymmetrically patterned catalytic Janus colloids in the seminal work by Anderson~\cite{Anderson1989}. 

Catalytic Janus colloids produce rapid motion in fluids by decomposing fuel asymmetrically around their body and taking advantage of the resulting gradients to develop motion through the corresponding phoretic flows. 
In principle, the asymmetric patterning of the colloidal particle can lead to a differential drag that produces a torque when the motion of the particle is not aligned with the polar axis of symmetry of the Janus pattern. 

Considering a ‘slip–stick’ spherical particle whose surface is partitioned into slip and no-slip regions, the coupling between torque and translation, as well as between force and rotation, can be obtained in the form of a Faxen-type formula in the limit where the slip length is small compared to the size of the particle~\cite{Swan2008, Premlata2021}. 
This coupling,  which is uncharacteristic of spherical particles in unbounded Stokes flow, originates purely from the slip–stick asymmetry, and generically produces self-alignement. 
It is however likely that the difference of slip length between the two sides of a real Janus microswimmer is so small that, in practice, the coupling coefficient is also very small. Nevertheless, the alignment with external forces was recently used to explain the chiral motion observed in light activated Janus colloids (spherical silica particles half-coated with carbon) that are moving in a viscoelastic fluid~\cite{Narinder2018}.

\subsubsection{Migrating cells}
\label{subsubsectionCells}
There is experimental evidence that epithelial-type cells on a substrate exhibit self-alignment properties at the collective level. 
Specifically, two of the main signatures of self-alignment, self-organized flocking in open boundary conditions~\cite{Szabo2006,Malinverno2017} and oscillations in confinement~\cite{Deforet2014,Petrolli2019,Peyret2019}, have been observed in epithelial cell sheets. An explicit biological role for self-alignment in the complex migration and deformation behaviour of the simple sheet-like animal Trichoplax has also recently been suggested \cite{Bull2021,davidescu2023growth}.

Unlike in the case of mechanical agents, studying the active mechanics of cells in isolation is difficult, and emergent dynamical equations are only beginning to be proposed \cite{Brueckner2022}. There is however consensus that, at the single cell level, a process called Contact Inhibition of Locomotion \cite{Abercrombie1954, Smeets2016, Stramer2017} slows and reorients cells when they encounter an obstacle.
Single-cell dynamics is also of limited use to their function in a tissue due to cells transitioning from a mesenchymal migratory state (e.g the keratocytes in~\cite{Szabo2006}) to an densely packed epithelial state with very different morphology. The latter is dominated by strong cell-cell junctions that incorporate both attractive-repulsive forces and active forces between cells, in addition to the individual active migration over the substrate \cite{alert2020physical}. 
Cell polarisation as an internal state (called planar cell polarisation in the biology community) is then variously defined as cell elongation, cell migration direction, actomyosin cortex polarisation, or an anisotropy in chemical expression, with unclear distinctions. A number of competing feedback mechanisms related to activity have been identified in cells, that it is imperative to disentangle. Separately, the plithotaxis mechanism \cite{Tambe2011} makes cells migrate along the direction of principal stress.
In the active matter community, hydrodynamic theories such as the active gel theory \cite{Juelicher2018} integrate this internal state information into a polarisation vector, which in active nematics corresponds to the nematic orientation tensor, while motion is due to gradients in the stresses generated by the polarisation. 
%The hallmark of active nematic states are the appearance of motile $\pm 1/2$ Topological defects, which have been observed, on average, in epithelial cell sheets \cite{saw2017Topological}. 
However, the connection between these stress gradients and the \emph{individual} migration and feedback mechanisms is unclear. 

Capturing these types of feedback has led to several computational cell models that include self-alignment, where cells are represented by network connextions \cite{davidescu2023growth}, particles \cite{Szabo2006,Smeets2016}, vertex models \cite{Barton2017,Petrolli2019,Malinverno2017,Giavazzi2018}, or phase field models \cite{Peyret2019,monfared2023mechanical}. We note that, given the strongly nonlinear internal dynamics of cells in response to external mechanical perturbations, the normalized $\boldsymbol{w} = \hvb$ velocity coupling has generally been used in these models.
%\vd{\bf [This parapraph is too long, but I do not see where to cut it]}
%\ch{[Maybe now it's ok?]}

\section{Collective motion of self-aligning agents}
\label{sec:collmot}
% ====================================
Collective motion is the hallmark of mutually aligning active agents.
The field of active matter actually emerged with the introduction of the Vicsek model~\cite{Vicsek1995} to study the emergence of large scale collective dynamics in an out of equilibrium system inspired by bird flocks. As stated by the authors, ``the only rule of the model is that at each time step a given particle driven with a constant absolute velocity assumes the average direction of motion of the particles in its neighborhood of radius $r$ with some random perturbation added". It is now clearly established that this model exhibits a discontinuous transition, from a disordered gas to a polar liquid phase, with true long-range orientational order~\cite{Toner2005,Chate2008,Peshkov2014,Solon2015}.  
It is important to realize that the Vicsek model does not specify the origin of the mutual alignment amongst its agents. As such, it must be seen as an effective model, the strength of which is to potentially encompass a large class of systems~\cite{Vicsek2012}, from schools of fish~\cite{Niwa1994}, herds of quadrupeds~\cite{ginelli2015intermittent}, flocks of flying birds~\cite{Cavagna2014}, or bacterial colonies~\cite{Zhang2010}, to actin filaments~\cite{Schaller2010}, vibrated polar discs~\cite{Deseigne2010}, or rolling colloids~\cite{Bricard2013}. 

Whether of physical or social origin, the mutual alignment of the velocity is most often seen as the consequence of pairwise interactions that directly couple the orientational degrees of freedom of the agents. For instance, self-propelled rods~\cite{Peruani2006,Ginelli2010,Peruani2011} tend to align because physical steric repulsion forces induce an explicit pairwise torque on their bodies, while birds are thought to align because they try to match their neighbors' velocities, due to ``social forces''.
However, this is not always the case: there are various models of self-propelled agents that produce collective motion and have interactions that depend on relative positions instead of relative angles, since they require no explicitly aligning interaction. For example, collective motion can be driven by escape-pursuit dynamics~\cite{Romanczuk2009}, by the conservation of momentum in inelastic collisions of self-propelled isotropic particles~\cite{Grossman2008}; or by the deformation of self-propelled soft particles with local repulsion~\cite{Menzel2012}.

In this context, self-alignment models form a class of their own, with a distinct coupling between the translational and orientational degrees of freedom at the level of a \emph{single} particle. We shall nevertheless see below that it also leads to collective motion under various circumstances. 

\subsection{Liquids of self-aligning agents}
\label{sec:sa_liquids}
\begin{figure}[t!]
\center
\includegraphics[width=0.95\columnwidth]{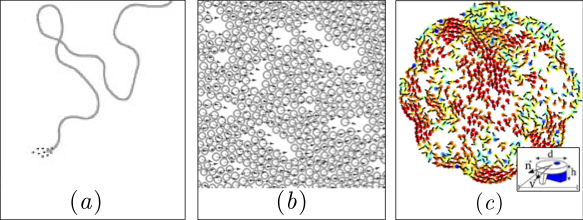}
\caption{{\bf Collective motion in self-aligning liquids:} 
(a) Early simulations in the first model of self-aligning agents (adapted from~\cite{Shimoyama1996}); 
(b) simulations in a large population of self-propelled self-aligning disks (adapted from~\cite{Szabo2006}; 
(c) experimental evidence of collective motion in a system of vibrated polar, disk-shaped grains (adapted from~\cite{Deseigne2010}).}
\label{fig:collmot_phenomenology}
\end{figure}

To our knowledge, the very first work where self-alignment was introduced dates back to the early years of active matter~\cite{Shimoyama1996}. Interestingly, it was introduced in a model specifically designed to describe collective motion. The central ingredient of the model is to take into account the heading unit vector $\hnb$ and realize that ``In a glide, the heading and the velocity vector need not be parallel". The authors therefore assume that the heading relaxes to the direction of the velocity in a finite time. The dynamics of agent $i$ was hence described by the following noiseless equations:
\bseq \label{eq:Shimoyama}
\begin{align}
	m \ddot\rb_i &= F_a \hnb_i - \gamma \dot\rb_i  +  \sum\limits_{j\neq i} \alpha_{ij}\fb_{ij} + \gb_i, \label{eq1:Shimoyama} \\
	\tau_a \dot{\hnb}_i &= (\hnb_i \times \hvb_i) \times \hnb_i, \label{eq2:Shimoyama}
\end{align}
\eseq
corresponding to Eq.~(\ref{eq:New_aod}), in the case with $\wb = \hvb$ and an isotropic translational damping. The forces acting on the particles are the pairwise interaction forces $\alpha_{ij}\fb_{ij}$, which in this case are not necessarily isotropic, and a global cohesive force $\gb_i$. The interaction forces and this global cohesive force share the same amplitude $c$.
After rescaling length by the interaction range $r_c$ and time by $\tilde{t}_0 = r_c/v_a$, where $v_a = F_a/\gamma$ is the steady state active speed of a free agent, the authors introduce three dimensionless parameters: $P=\frac{\gamma r_c}{\tau_a F_a}$,
$R=\frac{mF_a}{\gamma^2 r_c}$, $Q=\frac{F_a}{c}$. Here, $P$ is the ratio between the interaction time $r_c \gamma/F_a$ and the alignment time $\tau_a$ of the orientation; $R$ is the inertia to damping ratio; and $Q$ is the active force to interaction force ratio. 
The authors consider the overdamped regime where $R\ll1$ and report a transition from disordered dynamics to collective motion controlled by the ratio $G=P/Q$: at large $G$ the dynamics are chaotic and analogous to the swarming of mosquitos; at small $G$ the dynamics are highly polarized and analogous to the marching of crane. In the transitional regime the collective motion is wandering, in a way akin to the motion of sparrows (Fig.~\ref{fig:collmot_phenomenology}-a). The influence of noise is not reported.

Ten years has to pass before a similar model including self-alignment was introduced in~\cite{Szabo2006} to describe the collective migration of tissue cells. Here, the $i$-th cell's position $\rb_i$ and orientation $\hnb_i = (\cos\theta_i, \sin\theta_i)$ evolve in a 2d plane following the overdamped equation of motion
\bseq \label{eq:Szabo}
\begin{align}
	\dot\rb_i &= v_0 \hnb_i + \alpha  \sum_{j=1}^{N} \Fb_{ij}, \label{eq1:Szabo} \\
	\dot{\theta}_i &= \frac{1}{\tau_a} \sin^{-1}\left[\left(\hat{n}_i \times \hat\vb_i\right)\cdot \hat{e}_z\right] + \sqrt{2D_\theta}\,\eta_i, \label{eq2:Szabo}
\end{align}
\eseq
where $\hat{e}_z$ is a unit vector orthogonal to the plane of motion and $\eta_i$ is a random variable that introduces delta correlated Gaussian white noise with zero mean, \ie with $\langle\eta_i(t)\rangle=0$ and $\langle\eta_i(t)\eta_j(t^{\prime})\rangle=\delta_{ij}\delta(t-t^{\prime})$.
The first equation is the overdamped version of Eq.~(\ref{eq:New_trans}), with $v_0 = F_a/\gamma$ when the damping is isotropic. The second equation can be recast in a simpler form:
\beq
	\dot{\theta}_i = \frac{1}{\tau_a} (\psi_i-\theta_i) + \sqrt{2D_\theta}\,\eta_i,
\eeq
where $\psi_i$ denotes the orientation of the velocity, $\hvb_i = (\cos\psi_i, \sin\psi_i)$. It thus amounts to a linearization of the overdamped version of Eq.~(\ref{eq:New_rot_od}), in the $\wb = \hvb$ case.

In contrast to the model proposed by Shimoyama \textit{et al.}, this model incorporates angular noise but does not include long-range global forces. The resulting dynamics exhibit a continuous transition from a disordered to an ordered state, which can be achieved by reducing noise or increasing density (Fig.~\ref{fig:collmot_phenomenology}-b). The model was also used to examine the impact of confinement, revealing the emergence of circular motion across a wide range of parameters, which is consistent with experimental observations of cellular behavior.

Self-alignment was first identified experimentally when it was found to be the root cause of the onset of the collective motion observed in a system of vibrated polar discs~\cite{Deseigne2010,Deseigne2012}. In this setup, millimeter-size objects were designed to be self-propelled agents that advance along a well-defined axis when placed on a vibrating surface, while remaining disc-shaped with respect to collisions and thus interacting with neighbors through central, isotropic forces (see Fig.~\ref{fig:collmot_phenomenology}-c). 
In~\cite{Weber2013}, the authors developed a faithful model for the motion, collisions, and self-alignment of such polar disks, in order to extend their experimental observations to larger in silico systems. 
The model they introduced reads:
\bseq \label{eq:Weber}
\begin{align}
	m \ddot\rb_i &= F_a \hnb_i - \gamma \dot\rb_i  +  \sum\limits_{j\neq i} \fb_{ij} + \eta^\parallel_i  \hnb_i  + 	\eta^\perp_i \hnb_i^{\perp} \label{eq1:Weber} \\ 
	\dot\theta_i &= \beta\, {\rm sign}\left[\cos(\psi_i-\theta_i)\right]\, \sin(\psi_i-\theta_i), \label{eq2:Weber}
\end{align}
\eseq
with a self-alignment of $\hnb$ towards $\hvb$. The agents interact via inelastic collisions encoded in the pairwise forces $\fb{ij}$. The damping is isotropic, but the ``active noise'' is anisotropic, respecting the particle's polar symmetry: the random variables $\eta^{\parallel,\perp}_i$ follow a Gaussian distributed white noise with zero mean, i.e.  $\langle \eta^{\parallel,\perp}_i(t) \eta^{\parallel,\perp}_j(t') \rangle= 2D_{\parallel,\perp} \delta_{ij} \delta_{{\parallel \perp}} \delta(t-t')$, where $D_{\parallel,\perp}$ is the corresponding diffusion constant. There is no noise on the angular dynamics.
One interesting specificity of this model is that the sign of the coupling changes according to $\alpha_i = \angle(\vb_i, \hnb_i) = \theta_i - \psi_i$, the angle between velocity and polarity. For $|\alpha_i|>\pi/2$, it was assumed that frictional interactions with the vibrating plate would rotate $\hnb_i$ towards $\vb_i$, producing self-alignment, while for $|\alpha_i|>\pi/2$, it was assumed that $\hnb_i$ would instead rotate towards $-\vb_i$. 

By calibrating the model parameters using experimental data from the single-particle dynamics and collision statistics, the authors were able to obtain quantitative agreement between the experiments and the model also at the collective level. From there, they could show firm numerical proof of the transition to collective motion in this system. 
Given the isotropic pairwise interactions of the disks, this result underlines that collective motion can emerge from self-alignment alone. It is thus tempting to think of self-alignment as one specific microscopic mechanism (amongst others) that can lead to a form of effective pairwise mutual alignment, which would thereby belong to the class of active matter systems described by the Vicsek model~\cite{Vicsek1995}. 

\begin{figure}[t]
\vspace{-0mm}
\center
\includegraphics[width=0.95\columnwidth,trim = 0mm 0mm 0mm 0mm, clip]{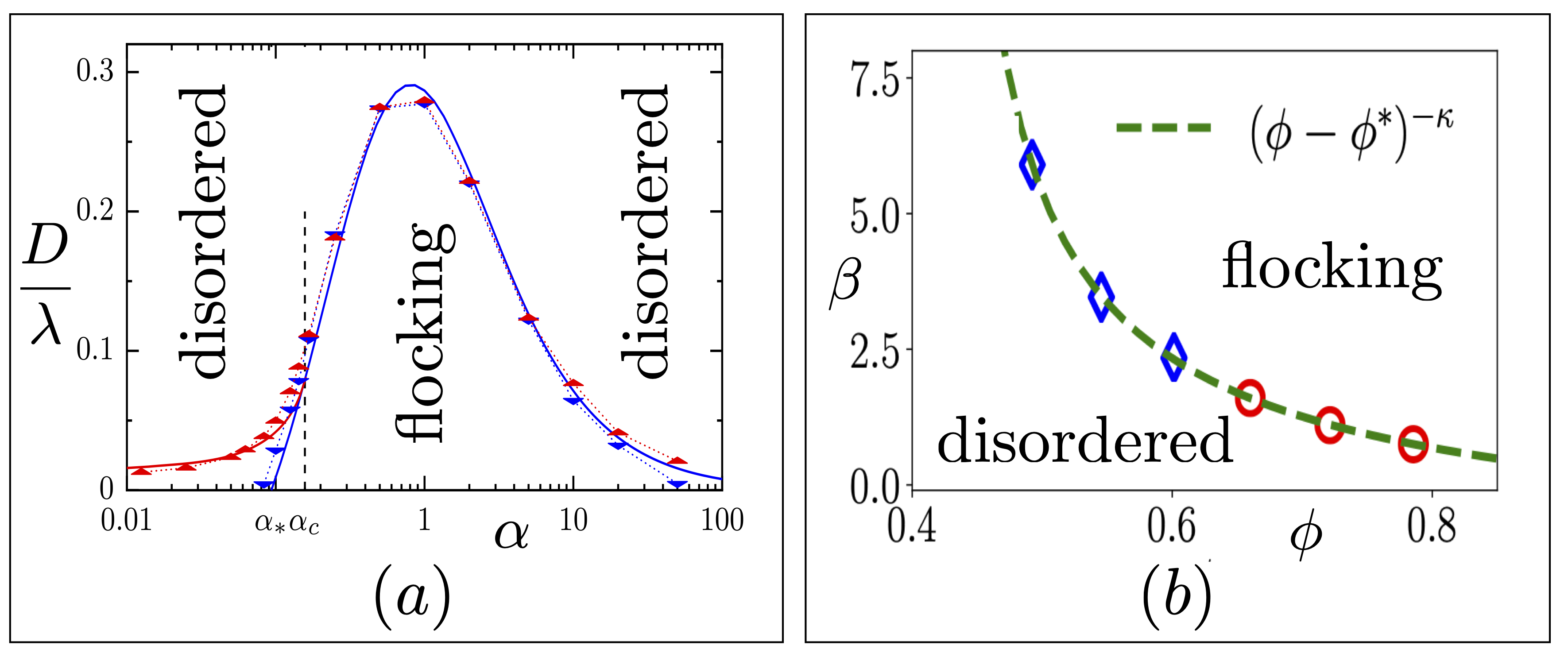}
\vspace{-0mm}
\caption{{\bf Phase diagrams for collective motion in liquids composed of self-aligning particles:} (a) as obtained numerically (blue and red data points) from the simulations of equations~\ref{eq:Lam} with density $\rho=10^{-2}$ and analytically in the limit of low densities and $\tilde\tau_v\rightarrow\infty$. $D/\lambda$ denotes the ratio of the angular noise amplitude to the collision rate and $\alpha = \tilde\tau_n / \tilde\tau_v$ (adapted from~\cite{Lam2015});
(b) as obtained numerically from the simulations of equations~\ref{eq:Szabo}; at large packing fractions (red markers), the transition is continuous; at small packing fraction (blue markers), it is discontinuous (adapted from~\cite{Paoluzzi2022}).
}
\label{fig:collmot-liq-phasediag}
\vspace{0mm}
\end{figure}

However, the situation is not that simple, as demonstrated in~\cite{Lam2015a,Lam2015}, where the Landau terms of the large scale hydrodynamics were derived for a system of self-propelled hard-disks obeying self-alignment, in the $\wb=\vb$ case. The authors adopt the same units and time scales as~\cite{Shimoyama1996}, with $r_c$ being the diameter of the hard disks, and write the dimensionless equation of motion as 
\bseq \label{eq:Lam}
\begin{align}
	\tilde\tau_v \ddot\rb &= \hnb - \dot\rb  +  \sum\limits_{j\neq i} \fb_{ij}, \\
	\tilde\tau_n \dot{\hnb} &= (\hnb \times \dot\rb) \times \hnb + \sqrt{2\Dr} \eta\,\hnb^{\perp}, \label{eq:sa_rot_noise}
\end{align}
\eseq
with $\tilde\tau_v = \frac{m F_a}{\gamma^2 r_c}$ and $\tilde\tau_n = \frac{1}{\beta r_c}=\frac{l_a}{r_c}$. 
When two particles obeying the above dynamics collide, even elastically, the polarities of the particles and their respective velocities strongly misalign. This leads to a relaxation process during which the polarity and the velocity rotate toward each other until $\vb_i = \hnb_i$. Each collision followed by this relaxation process was named a ``scattering event''. For symmetric collisions, it is easy to see that such scattering event will induce an alignment of the velocities of the two particles. 

Using these equations and integrating over all possible pairwise events, the authors computed the effective total pairwise alignment strength as a function of the incoming angle of the collisions and thereby obtained the mean field phase diagram of the model in the low density limit (see Fig.~\ref{fig:collmot-liq-phasediag}-a). In the absence of noise, a strongly discontinuous transition from the disordered state to a highly polar liquid, or the flocking state, takes place when the ratio $\alpha = \tilde\tau_n / \tilde\tau_v$ exceeds a critical value $\alpha^*(\rho)$ which decreases with the density $\rho$. If we add noise, the transition shifts to higher values of $\alpha$ and, for large enough noise, it becomes continuous.
For a finite value of the noise, there is a reentrant transition towards the disordered state when $\alpha$ becomes too large. This can be understood intuitively as follows. When $\alpha$ is small, the relaxation of the polarization towards the velocity is so fast that self-alignment does not affect the orientation of the velocity and the liquid remains disordered. When $\alpha$ is too large, the persistence of $\hnb$ is so large that it is not influenced by the collisions with other agents; mutual alignment cannot set in and the liquid remains disordered. 
The phase diagram displayed on Fig.~\ref{fig:collmot-liq-phasediag}(a) is obtained in the limit of large $\tilde\tau_v$. Decreasing $\tilde\tau_v$, the left boundary of the flocking region remains unchanged, while the right one shifts towards larger $\alpha$. 

An important point made by the authors is that the dependence of the effective alignment strength on the incoming angle of the collision is markedly different from the one obtained from the Vicsek aligning rules. A qualitative consequence of this difference is that the mean field transition to collective motion is second order in the case of the Vicsek model, while we saw that its order depends on the amplitude of the noise in the case of self-aligning hard disks. Whether this qualitative difference observed at the mean field level has significant impact on the transition with spatial fluctuations remains an open issue, that is notoriously hard to address because the nucleation process leading to collective motion takes place at a bimodal far away from the spinodal.

Except for the presence of cohesion in~\cite{Shimoyama1996}, the deterministic dynamics introduced in this work and in~\cite{Lam2015} are essentially the same. Their result are complementary in the sense that they explore two different limits. Indeed, if we compare Eqs.~(\ref{eq:Shimoyama}) and (\ref{eq:Lam}), it is easy to see that $\tilde\tau_v=R$ and $\tilde\tau_n=P^{-1}$. While~\cite{Shimoyama1996} considers the overdamped limit $R\rightarrow 0$, ~\cite{Lam2015} studies the hard disk limit $Q\rightarrow 0$.
Together they establish that, in the absence of noise, the mean field transition to collective motion takes place for $\tilde\tau_n>\tilde\tau_n^*(\tilde\tau_v,Q)$ with  
\bseq
\begin{align}
&\tilde\tau_n^*(\tilde\tau_v\rightarrow 0,Q) \sim Q^{-1} \text{~\cite{Shimoyama1996}}\\
&\tilde\tau_n^*(\tilde\tau_v,Q\rightarrow 0) \sim \tilde\tau_v \,\,\,\,\,\,\text{~\cite{Lam2015}}
\end{align}
\eseq

Most models of self-propelled particles are considered in the overdamped limit, $\tilde\tau_v\rightarrow 0$, and then also consider softer potential than the strictly hard disk interaction considered in~\cite{Lam2015}. For such overdamped models, we will have $\vb_i = \hnb_i$ and no self-aligning term at all times except for the duration $\tau_i$ of the interaction, during which the self-alignment will reorient the $\vb_i$ velocities. According to the analysis of~\cite{Lam2015a}, what matters is the persistence of the polarization throughout the scattering event. It is thus the ratio $\tilde\tau_n / \tau_i$ which plays the role of the above parameter $\alpha$. The transition reported in~\cite{Szabo2006} is mean field like because of the small system size that are considered. Its continuous nature can thus be understood in light of the above discussion, as a transition taking place for large enough $\tilde\tau_n/\tau_i$.

When the density becomes so large that crowding effects become significant, equilibrium liquids experience a strong increase of their viscosity, eventually reaching the glass transition when crystallization is avoided. The interplay of activity and glassiness has recently attracted significant attention~\cite{Janssen2019}. However, most studies have not considered the presence of alignment, and even less the presence of self-alignment, except for the recent work in~\cite{Paoluzzi2022}, which studies a polydisperse system of particles with self-aligning dynamics described by
\bseq
\begin{align}
\dot\rb_i &= v_0 \hnb_i + \mu \Fb_i, \\
\dot\theta_i &= \beta \sin(\psi_i - \theta_i) + \sqrt{2D_r} \eta_i,
\label{eq:Paoluzzi}
\end{align}
\eseq
where $\psi_i$ is the direction of $\hvb_i$. The dynamics in this case is overdamped, as in~\cite{Szabo2006}, but the self-aligning term is not linearized in $\psi_i - \theta_i$.

The authors report that a transition to collective motion takes place when $\beta$ exceeds a density dependant threshold $\beta^*(\phi)$). Within the limit of their finite size simulations, they observe that the transition is first order at moderate values of $\phi$, while it becomes second order for large $\phi$ (Fig.~\ref{fig:collmot-liq-phasediag}-b)  and remains second order when increasing system size. 

For smaller self-alignment the authors observed the well known motility induced phase separation (MIPS), which takes place at large enough densities in systems of non aligning repulsive active Brownian particles~\cite{Cates2015}. This transitions appears to be mutually exclusive with the transition to collective motion. 
At higher densities, the system reaches instead a glassy state where the relative motion of particles ceases. Nevertheless, the flocking transition is not suppressed in this regime, and the authors distinguish ``glassy'' from ``flocking glass'' states. 

\subsection{Collective motion in a confining potential}
\label{sec:CollmotParabola}
Very recently, the collective dynamics of self-aligning polar active matter, interacting repulsively through a 
truncated Lennard-Jones (WCA) potential, and confined in a harmonic potential, have been investigated numerically in~\cite{canavello2023polar}. Here, the particles obey the overdamped dynamics described by Eq.~(\ref{eq:New_od}) in the case with $\wb = \vb$.

\begin{figure}[t]
\vspace{-0mm}
\center
\includegraphics[width=0.95\columnwidth,trim = 0mm 0mm 0mm 0mm, clip]{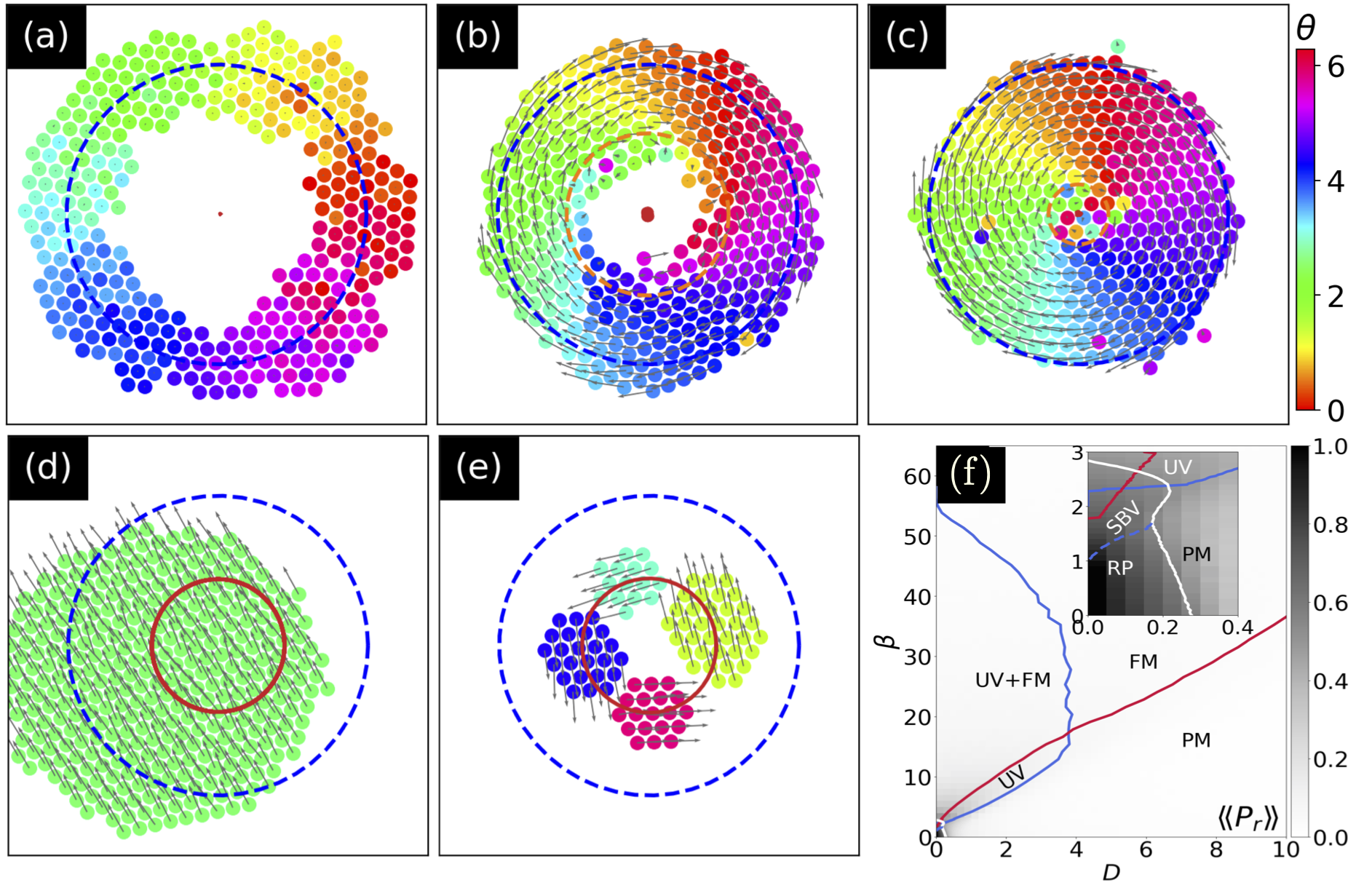}
\vspace{-0mm}
\caption{{\bf Collective motion in an harmonic potential:} (a-e) The main dynamical states observed, when varying the self-aligning strength $\beta$ and the rotational noise $D$, with the color coding for the orientation of the particle: (a) the radially polarized (RP) state, (b) the shear banded vortex state (SBV), (c) the uniform vortex state (UV), (d) the orbiting ferromagnetic state (FM) and (e) the multi orbiting polar clusters state observed at lower packing fraction. (f) Phase diagram, with the gray intensity coding for the radial polarization; inset: zoom on the small $\beta$, small $D$ region; (adapted from~\cite{canavello2023polar}).
}
\label{fig:collmot-parabola}
\vspace{0mm}
\end{figure}

The main output of this work is a phase diagram and a careful description of the dynamical phases, as summarized in Fig.~\ref{fig:collmot-parabola}. 
In the small $\beta$ and large noise $D$ region, all polar order parameters vanish, which characterizes the unpolarized or paramagnetic (PM) state.  Decreasing $D$ while maintaining a small $\beta$, here typically smaller than one, the dynamics enters the radially polarized (RP) state displayed in Fig.~\ref{fig:collmot-parabola}-a, where all particles point outward. Increasing $\beta$ while keeping $D$ small, the system organizes into a vortex state, first in a shear banded vortex (SBV) state shown in Fig.~\ref{fig:collmot-parabola}-b, which rapidly turns into the uniform vortex (UV) state in Fig.~\ref{fig:collmot-parabola}-c for large enough $\beta$.
This uniform vortex does not rotate as a strict rigid body but rather as a deformable solid. Increasing $\beta$ at large noise, the system adopts a ferromagnetic state, where a polarized cluster revolves without rotation (FM, Fig.~\ref{fig:collmot-parabola}-d). When the packing fraction is low, this unique cluster may break into several smaller clusters (FM, Fig.~\ref{fig:collmot-parabola}-e). 

The authors also report that the orbiting ferromagnetic phase is highly resilient to noise, being stable even at high values of $D$, specially for high angular mobility, while the uniform vortex phase is typically much more sensitive to fluctuations. As shown in the phase diagram (Fig.~\ref{fig:collmot-parabola}-f), the uniform vortex and the ferromagnetic states coexist for low enough noise. As a result, the transition from the uniform vortex state to the ferromagnetic one and back exhibits pronounced a hysteretic behavior. 

% AQUI 6
\subsection{Collective motion in active Voronoi models}
\label{sec:ActiveVoronoi}
\begin{figure}[t]
\vspace{-0mm}
\center
\includegraphics[width=0.95\columnwidth,trim = 0mm 0mm 0mm 0mm, clip]{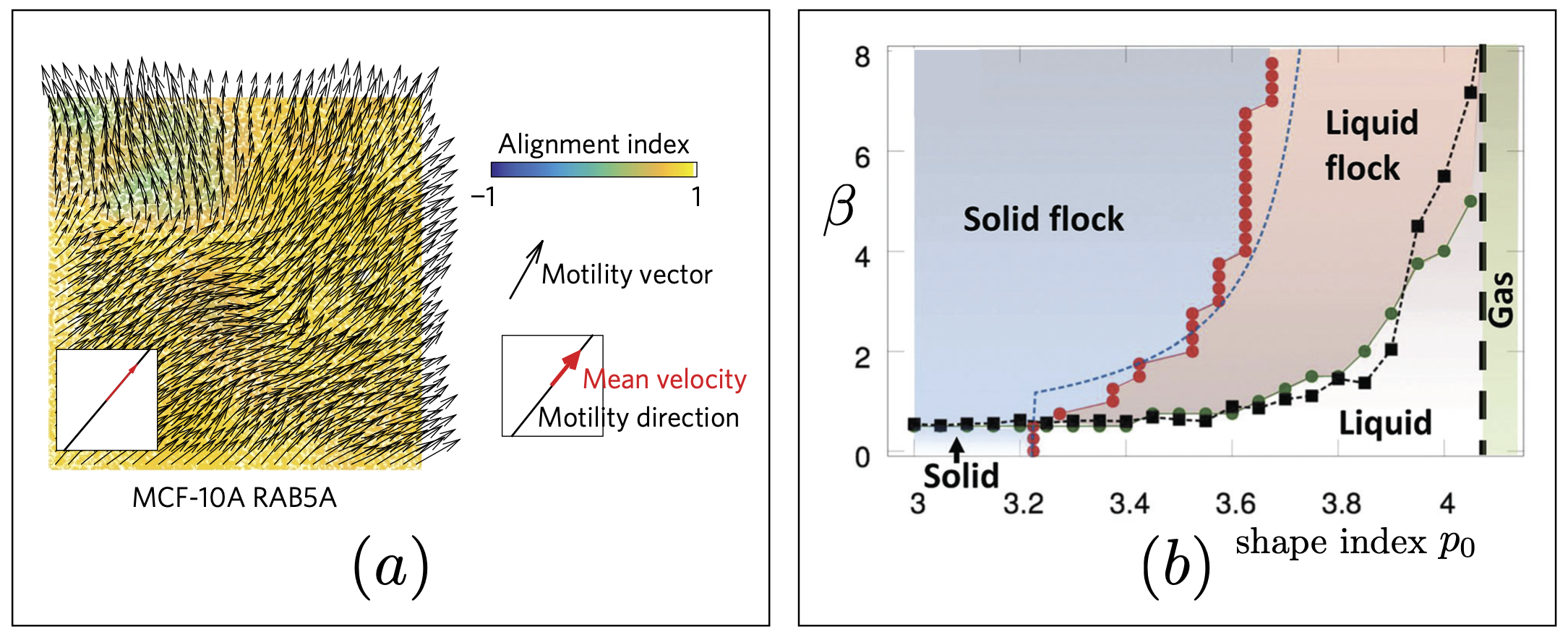}
\vspace{-0mm}
\caption{{\bf Epithelial tissues and Active Voronoi models:} (a) Human mammary epithelial cells treated with RAB5A-MCF-10A, inducing flocking motion, (adapted from \cite{Malinverno2017}); (b) Alignment-shape index phase diagram as obtained from a self-propelled Voronoi model (adapted from~\cite{Giavazzi2018}).
}
\label{fig:tissue-voronoi}
\vspace{0mm}
\end{figure}

Epithelial cell sheets are confluent monolayers of active cells where individual cells interact both with a substrate and with other cells in an arrangement that resembles cobblestone pavement (Fig.~\ref{fig:tissue-voronoi}-a). To model this geometry, the Vertex model of epithelial cell sheets \cite{Nagai2001,Farhadifar2007} maps them to a planar polygonal tiling in which cell shapes were, initially, determined by energy minimisation as in foam models (see \cite{Fletcher2014} for a review). 
In the now consensus energy functional
\begin{equation} 
V_{\text{vertex}} = \sum_{i=1}^N \frac{\kappa}{2} (A_i - A_0)^2 + \frac{\Gamma}{2} (P_i-P_0)^2, 
\end{equation}
the shape of each cell is constrained by an area stiffness $\kappa$ (that constrains fluctuations away from a target area $A_0$) and a perimeter stiffness $\Gamma$ (that likewise constrains perimeter fluctuations). This energy is expressed as a function of the \emph{vertex positions} $\rb_{\mu}$ of the cell polygonal shapes.

To make the model above active, \cite{Bi2016} incorporated cell crawling motility over a substrate by introducing overdamped Langevin dynamics akin to active Brownian motion,
\begin{align*} 
\dot{\rb}_i &= v_0 \hnb_i - \alpha \bm{\nabla}_i V_{\text{vertex}}, \\
\dot{\theta}_i &= \sqrt{2D_\theta} \eta_i. 
\end{align*}
While conceptually straightforward, the presence of the gradient with respect to the cell \emph{centre} positions $\rb_i$ requires a one-to-one continuously differentiable map between these and the vertex positions $\rb_{\mu}$. This can be solved by using the dual between a Delaunay triangulation among cell centres and a Voronoi tiling for the vertex positions, which led to the adoption of the name self-propelled Voronoi (SPV) model. The SPV model admits two phase transitions between a rigid solid state and a liquid state (Fig.~\ref{fig:tissue-voronoi}-b): one when cell motility $v_0$ becomes large enough; the other when cells have a more elongated shape, as quantified by the shape index $p_0 = \frac{\sqrt{P_0}}{A_0}$. This index measures the preferred perimeter relative to the preferred area, with the system being liquid (for $v_0=0$) if $p_0$ is above a critical $p_0^*\approx 3.81$~\cite{Park2015}.

In the SPV, self-alignment can be added as an explicit alignment torque using the same mechanism discussed before \cite{Barton2017},
\begin{align} 
&\dot{\rb} = v_0 \hnb_i - \alpha \bm{\nabla}_i V_{\text{vertex}}, \\
&\gamma_r \dot{\theta}_i = \bm{\tau}_i \cdot \hat{\boldsymbol{z}} + \sqrt{2D_\theta} \eta_i, \quad \bm{\tau}_i = J_v (\hnb_i \times \hvb_i)\times \hnb_i, \label{eq:AJVoronoi}
\end{align}
where $\hat{\boldsymbol{z}}$ is the out of plane unit vector. This model produces a flocking state at sufficiently strong alignment $\beta=J_v/\gamma_r$. 
It has been used to investigate the flocking transition in cell sheets in \cite{Malinverno2017} and \cite{Giavazzi2018}.

More specifically, in \cite{Malinverno2017} the authors developed an experimental system of human mammary epithelial cells. In its natural state, this cell sheet acts as a liquid with spatiotemporally correlated motion, but without global flocking. Treating it with RAB5A-MCF-10A, which weakens the intercellular adhesive bonds, has been empirically shown to induce a transition to flocking motion in the sheet (Figure \ref{fig:tissue-voronoi}(a)). Including velocity self-alignment, as introduced above in the dynamics, recovers this transition, now as a function of $J_v$. 
Separately, in \cite{Giavazzi2018} the authors mapped the full phase space of this model as a function of the shape parameter $p_0$ and of the alignment strength $\beta=J_v/\gamma_r$ for constant active driving and rotational noise (Figure \ref{fig:tissue-voronoi}(b)). Similar to the particle-based case, this model admits both a solid to liquid transition and a flocking transition, including a solid flocking phase. The precise nature of these transitions and of the flocking phase have so far not been explored in detail.

\subsection{Collective motion in a crystal of self-aligning discs}
If we consider high densities, the kinetic theory calculation that demonstrates the emergence of collective motion in a dilute assembly of self-aligning hard discs~\cite{Lam2015a} is not supposed to hold. This is all the more true when the time separating two collisions becomes shorter than the relaxation times of the polarity and velocity. One would thus expect collective motion to be preempted by crowding effects at large densities. Surprisingly enough, this is not the case, as demonstrated in~\cite{Briand2016,Briand2018}, where the authors investigated both experimentally and numerically very dense assemblies of mono-disperse self-aligning hard discs. They found two emerging effects of the active self-aligning dynamics.

First, the crystallization transition obeys a scenario that is radically different from the equilibrium case. The transition toward the crystal phase is marked by the emergence of close-packed crystallites, which coexist with a surrounding moving fluid. Increasing the density, the crystallites merge into a large hexatic phase populated with strongly dynamical defects, leading to a complete decoupling of the dynamics and the structure. While the structure is dominated by that of a closed packed crystal, the mean square displacement exhibits no plateau, and remain super-diffusive on long time scales. There is no truly slow dynamical regime and all particles rearrange their position with respect to their neighbours on a rather modest timescale.

\begin{figure}[t]
\center
\includegraphics[width=0.95\columnwidth]{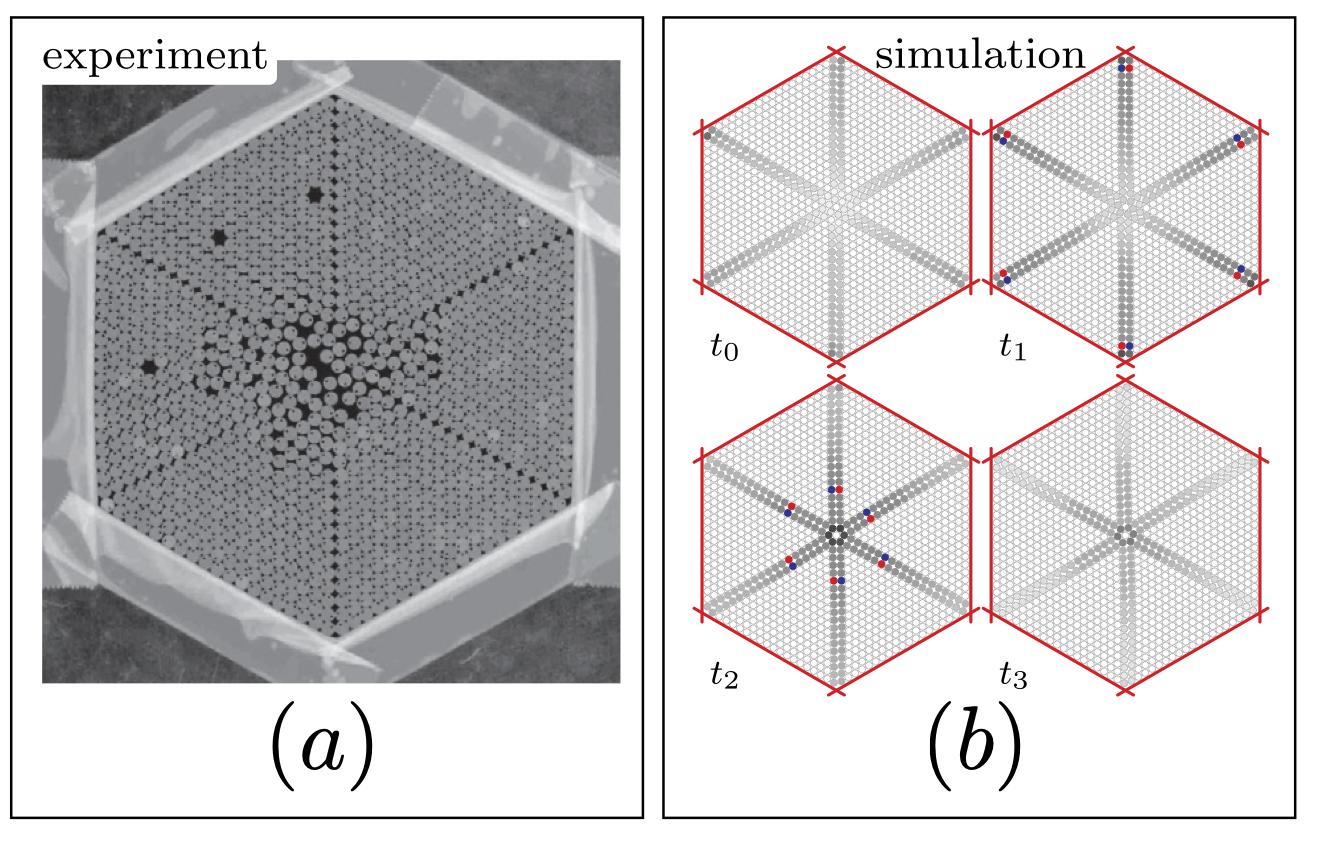}
\caption{
{\bf Collective motion in a crystal of self-aligning discs:} 
(a) A very dense packing of self-propelled vibrated disks (packing fraction $\phi = 0.86$) confined in a hexagonal arena forms a self-flowing crystal, with shear localization along default lines emerging from the corner of the hexagon and merging in its center (adapted from~\cite{Briand2018}); 
(b) Simulation of the same system, at very low noise and even higher packing fraction ($\phi = 0.88$), produce the same phenomenology with an even higher level of order, both in space and time (adapted from~\cite{Briand2018}).}
\label{fig:collmot_crystal}
\end{figure}

Second, and more surprisingly, the authors report the existence of a flowing crystalline phase.
After a long transient, a perfect crystalline lattice with initially random orientation of the polarities starts moving as a whole with all particles aligned in the same direction: despite the high frequency collisions, self-alignment still produces collective motion. When the crystal is prepared in a hexagonal arena that respects the same crystalline symmetry, the whole system spontaneously forms a macroscopic sheared flow, while conserving an overall crystalline structure (Fig.~\ref{fig:collmot_crystal}). This flowing crystalline structure, which was called a ``rheocrystal,'' is made possible by the condensation of shear along localized stacking faults. In the presence of noise, the core of the system concentrates the defects and remains disordered (Fig.~\ref{fig:collmot_crystal}-a). The reader will note the strong similarities between this flowing crystal and the uniform vortex state reported on Fig.~\ref{fig:collmot-parabola}-(c).
Performing simulations for larger system sizes, with experimentally realistic packing fraction and noise level, the disordered core was observed to occupy a smaller fraction of the system, while the flowing velocity slowly converged to a higher value.
Numerically it is also possible to reduce the noise and increase the packing fraction to a point where the structure is defect free in the core and the dynamics eventually becomes periodic in time (Fig.~\ref{fig:collmot_crystal}-b). The highest packing fraction for which the crystal flows is controlled by the fraction of space needed for the stacking faults to take place. The later being subextensive, the flowing crystal phase range enlarges with increasing system size. 

%AQUI 7
\subsection{Collective motion of active elastic sheets}
\label{ABES}

Motivated by experiments performed with wheeled robots~\cite{Ferrante2012,Brambilla2013}, an active elastic sheet model was first introduced in~\cite{Ferrante2013,Ferrante2013a} to describe groups self-propelled agents moving on an arena while linked by linear springlike attraction-repulsion interactions. These agents obey the following overdamped equations of motion
\bseq  
\label{eq:Huepe}
\begin{align}
	\dot\rb_i &= v_0 \hnb_i + \apara  \left[ \left( \Fb_i  + \tilde{D}_{\rm r} \hxib^{\rm r}_i \right) \cdot \hnb_i \right] \hnb_i, \label{Eq.CH1}\\
	 \dot{\theta}_i &= \beta_\theta \left[ \left( \Fb_i + \tilde{D}_{\rm r} \hxib^{\rm r}_i \right) \cdot \hnb_i^{\perp} \right] + \tilde{D}_{\theta}\, \eta_i. \label{Eq.CH2}
\end{align}
\eseq
Noise was introduced in the headings and in the virtual interaction forces (corresponding to \emph{actuation} and \emph{sensing} noise, in the context of robotics) by adding the random variables $\eta_i$ (a delta correlated scalar with zero mean Gaussian white noise distribution, as in previous sections) and $\hxib^{\rm r}_i$ (a delta correlated randomly oriented unit vector), with amplitudes $\tilde{D}_{\theta}$ and $\tilde{D}_{\rm r}$, respectively.
The total forces over agent $i$ are evaluated as 
$\Fb_i = \sum_{j\in S_i} (-k/l_{ij}) (|\rb_{ij}|-l_{ij}) \rb_{ij}/|\rb_{ij}| $,
where $k/l_{ij}$ is the spring constant, $\rb_{ij}= \rb_j -\rb_i$, and $l_{ij}$ is the equilibrium distance between interacting agents $i$ and $j$.
Note that the interactions are permanent in this setting, so the interaction network remains unchanged throughout the dynamics. By connecting nearest neighbors on a plane we thus define a structure that can be viewed as an active elastic sheet.

A unique feature of this model is that the  mobility tensor $\alpha=\gamma^{-1}$ is fully anisotropic, with zero mobility along $\hnb_i^{\perp}$ and $\apara$ mobility along $\hnb_i$.
It thus describes agents that cannot translate in the $\hnb_i^{\perp}$ direction, because this would correspond to the wheeled robots sliding sideways.
Correspondingly, as described in Section~\ref{subsubsectionOffcentered}, the angular dynamics does not strictly result from self-alignment,
%(as $\dot\rb_i$ and $\hnb_i$ are always aligned),
and is instead equivalent to torques introduced by off-centered elastic forces, where $\beta_\theta$ results from the combined effect of the lever arm and the rotational mobility.
Note, however, that this equivalence is not complete, since the forces $\Fb_i$ in this model are computed from center to center, an artifact introduced to avoid the need for sensing other robots' headings.

\begin{figure}[t!]
\center
\includegraphics[width=0.95\columnwidth]{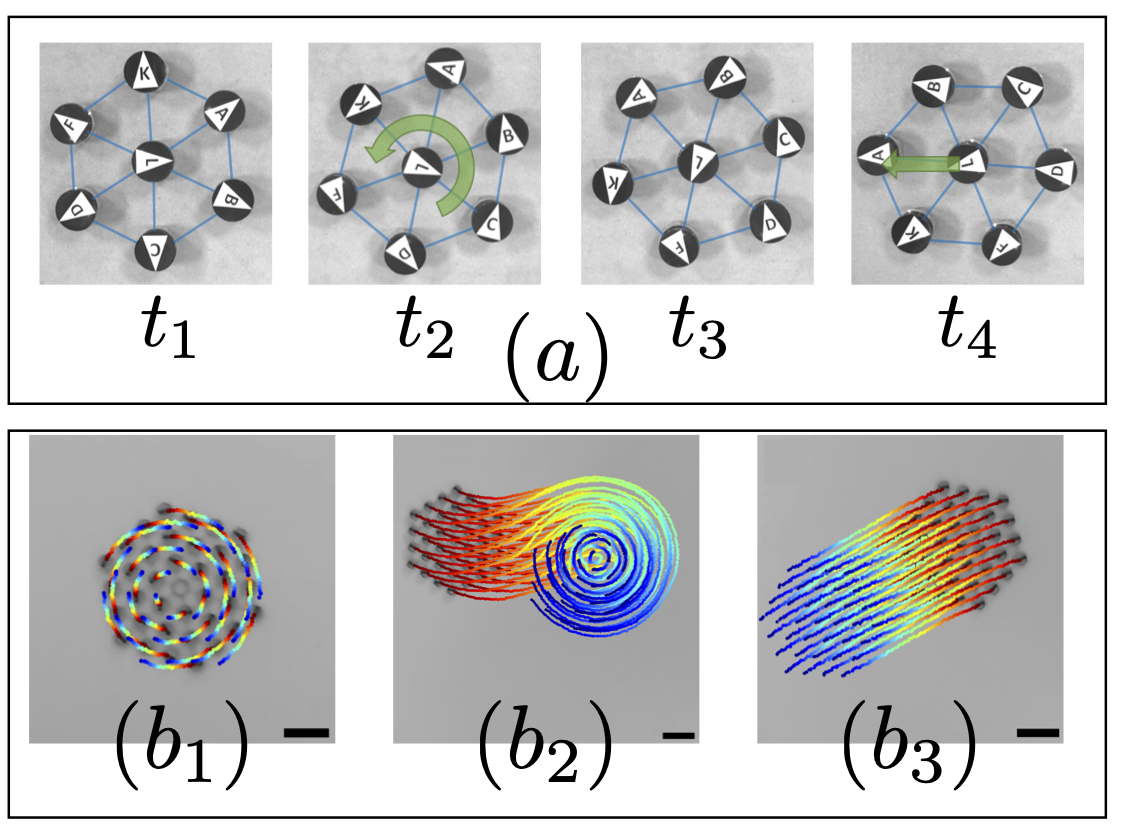}
\caption{{\bf Collective motion in active elastic sheets:} 
(a)  Successive snapshots of a robot swarm experiment that implements the position-based decentralized control algorithm based on equations~\ref{eq:Huepe}  and realizes translating collective motion ($t=t_4$), after sometimes visiting a metastable rotational state (here $t=t_2$); the overlaid blue lines indicate which robots are interacting (adapted from~\cite{Zheng2020});
(b) Active elastic structures obtained by physically connecting active units with springs (see section~\ref{sec:ActiveElasticLattice} and Fig.~\ref{fig:collact-hexbugs}-b) perform spontaneous solid body rotation ($b_1$) and translation ($b_3$), with the rotation being metastable with respect to translation ($b_2$) (adapted from~\cite{Hernandez2023}). 
}
\label{fig:collmot-AES}
\end{figure}

%CH: Discussion of nontrivial existence of long-range order - Apparent Mermin-Wagner violation
An elastic sheet of linked nearest neighbors following these dynamics self-organizes into a polarized state with common headings, despite having no explicit alignment forces~\cite{Ferrante2013a}. 
A numerical finite-size scaling analysis also showed long-range order at nonzero noise, in an apparent contradiction with the Mermin-Wagner theorem~\cite{Mermin1968} and in contrast to Vicsek-like models with fixed interaction networks~\cite{Vicsek2012}.
It was argued that this is because the self-organizing mechanism here is fundamentally different, based on the focusing of self-propulsion energy into lower elastic modes rather than on explicit local alignment.
%CH: Experiment realization using e-Pucks communicating with a computer
This model was implemented as a control algorithm in experiments with wheeled robots (e-Pucks) in~\cite{Zheng2020}, where it was shown that self-organized collective translation or rotation were reached for a broad range of parameters, despite multiple real-world limitations such as communication delays (Fig.~\ref{fig:collmot-AES}-a). However, these ordered states presented marginal linear stability and thus displayed persistent oscillations.

%CH: Emergence of scale-free correlations.
The same model was used to explore a potential alternative explanation for the scale-free correlations measured in bird flocks~\cite{Huepe2015}, which have been argued to result from their critical state~\cite{Cavagna2014}. Instead, it was suggested that the attraction-repulsion forces between birds produce an elastic-like coupling between active agents, exciting low elastic modes that trivially scale with the size of the system. A similar argument was also provided in an equilibrium model of self-aligning rotors~\cite{Casiulis2020}.

%CH: Complex interaction networks
In~\cite{Turgut2020}, the permanent nature of the interaction between agents in this model was exploited to compare the resilience of the aligned state to noise for nontrivial interaction networks, where elastic connections are established between agents beyond the nearest neighbors. The authors showed that, if we fix the mean number of interactions per agent, the critical noise will always increase as more random (long-range) links with Erdős–Rényi connectivity replace nearest neighbor interactions, but will instead reach a maximum and then decrease if these random links have a scale-free degree distribution~\cite{Newman2010}. Therefore, in contrast to most cases in network science, here the scale-free topology does not favor system integration.

The model was also recently extended~\cite{Lin2021} to describe a mechanical representation of an elastic sheet composed of self-propelled agents linked through linear springs attached to the tip of forward-projecting lever arms of length $R$. This version of the model uses the same Eqs.~(\ref{eq:Huepe}), but redefines the relative position with respect to which the forces are measured as the vector between the tips of the lever arms of the agents (where the physical springs would actually be attached), given by
$\rb_{ij} = (\rb_j + R \hnb_j)  - (\rb_i + R \hnb_i)$, instead of between the agent centers. 
For small $R/l_{ij}$, the interaction thus depends mainly on relative positions, not on relative angles, and the resulting states display the same features as in the original model.
For large $R/l_{ij}$, the interactions depend strongly on changes in the relative angles and become akin to the ferromagnetic-like alignment in the Vicsek model, displaying equivalent stationary states, including a loss of order for large enough systems at intermediate noise levels due to the Mermin-Wagner effect. 
For an intermediate range of parameters, the system can transition to a novel state of quenched disorder where agents display random fixed mean headings.

%AQUI 8
Finally, we note that the dynamics of a wheeled robot model with fully anisotropic mobility tensor, as described by equations~(\ref{eq:Huepe}), has to be inherently elastic. Indeed, the strong constraint imposed by the mobility tensor would make it a singular model in the limit where the active agents are connected by stiff springs. This singularity is removed when we relax this anisotropy and allow the agents to glide sideways. In this rigid limit, if we consider an isotropic mobility tensor, stress propagation alone induces self-organization, without exciting the vibrational modes. In the limit of rigid bonds, a mean-field-like transition is observed due to the truly long-range interaction between agents. In this limit, a general Landau like formalism can be established to understand the dynamics of collective motion along different floppy modes in such structures, from rigid body motion to folding mechanisms~\cite{Hernandez2023}.

\section{Collective actuation of active elastic materials}
\label{sec:collact}
Because self-alignment is not just effective pairwise alignment, but also alignment on the force field, one can expect new interesting dynamics to emerge when self-aligning polar agents are embedded in an elastic material attached to a frame. 
On one hand, the positional degrees of freedom of the active agents have a well-defined reference state. On the other hand, activity endows them with an additional degree of freedom in the form of polar active forces. These forces are expected to deform the elastic matrix and induce a stress strain field, which depend on the forces’ configuration, that is, on the agents' positions and orientations. The strain or stress tensor may, in turn, reorient the forces. This generic nonlinear elasto-active feedback opens the path towards spontaneous collective excitations of the solid, which were named collective actuation in~\cite{Baconnier2022}.
In the following, we shall start by describing collective actuation in a paradigmatic model experimental system composed of active elastic structures. We will then report the apparent  observation of collective actuation in jammed systems of particles and in active Voronoi models in confinement. Finally, we will discuss the observation of very similar dynamical behaviors in various biological systems.

\subsection{Selective and collective actuation in elastic lattices}
\label{sec:ActiveElasticLattice} 

\begin{figure}[t]
\includegraphics[width=0.95\columnwidth]{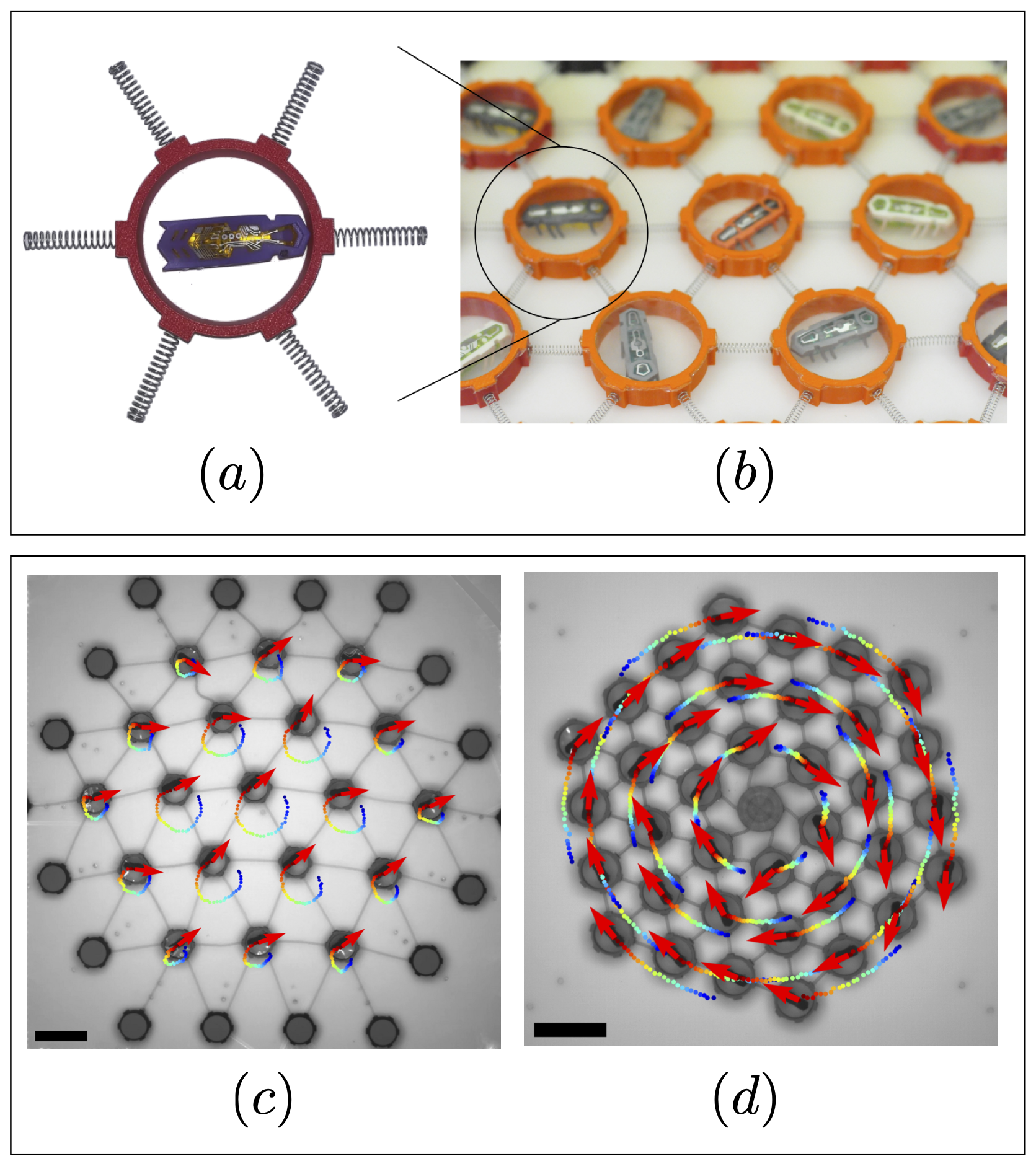}
\caption{\textbf{Collective actuation in model active elastic structures}. Active elastic building blocks, composed of (a) a hexbug inside a rigid annulus, are assembled with springs to form (b) an active elastic network (adapted from~\cite{Baconnier2022});
$(c)$ synchronized translational oscillations around the mechanical equilibrium configuration of a triangular lattice pinned at its edges (adapted from~\cite{Baconnier2022});
$(d)$ global alternating rotations of a triangular lattice with embedded central pinning; (adapted from~\cite{Baconnier2022}); the trajectories are color-coded from blue to red by increasing time; the red arrows indicate the polarities  $\boldsymbol{\hat{n}}_{i}$ at a given time; scale bars: $10$ cm.}
\label{fig:collact-hexbugs}
\end{figure}

A minimal experimental realization of an active elastic solid, with active polar units connected by springs (Fig. \ref{fig:collact-hexbugs}-a,b) was proposed in~\cite{Baconnier2022}. The authors focus on mechanically stable ordered lattices, where each node has a well-defined equilibrium reference position, but is displaced by the active agent. In contrast, each agent is free to rotate and to self-align with its displacement. This nonlinear feedback between deformation and polarization is characterized by two length scales: the typical elastic deformation caused by active forces $l_e$ and the self-alignment length $l_a$.
%AQUI 9

For small elasto-active coupling $\Pi=l_e/l_a$, the dynamics is disordered: displacements are small, and so is reorientation by motion, therefore angular noise dominates the polarity dynamics and randomizes the orientations. In contrast, for large enough  $\Pi$, collective actuation emerges. When the system is pinned at the edges, the collective actuation takes the form of synchronized chiral oscillations of the lattice nodes around their reference configuration (Fig. \ref{fig:collact-hexbugs}-c), here also breaking spontaneously the chiral symmetry present at the level of each agent. When the pinning takes place on a central node, both forbidding translation and free rotation, a global alternating rotation of the whole structure, analogous to the oscillations of a torsion-pendulum, emerges (Fig. \ref{fig:collact-hexbugs}-d). 
A closer examination of the dynamics reveals that, in both cases only a few normal modes are actuated, and crucially, they are not necessarily the lowest energy ones.

The dimensionless Newton equations describing this model  experimental system correspond to eq.~(\ref{eq:New_aod}), in the case where $\wb = \vb$, with the external force $\Fb_{\rm ext}$ given by the sum of the forces exerted by the elastic springs on each node. 
Using $l_0$, the spring rest length, and $t_0 = \gamma/k$ as length and time units, the dimensionless equations read:
\bseq \label{eq:Baconnier}
\begin{align}
\tau_{v} \ddot\rb_i &= \tilde{F}_{a}\hnb_i -  \dot\rb_i + \sum_{j \in \partial i} \left( |\rb_i - \rb_j| - 1 \right)\heb_{ij}  \\\tilde\tau_{n} \dot\hnb_i &= (\hnb_i \times \dot\rb_i) \times \hnb_i + \sqrt{2\tilde D} \xi \boldsymbol{\hat{n}}_{i}^{\perp}
\end{align}
\eseq
where  $\tau_v = \frac{m k}{\gamma^2}$, $\tilde\tau_n = \frac{1}{\beta l_0} = \frac{l_a}{l_0}$ and $\tilde F_a = \frac{F_a}{k l_0}=\frac{l_e}{l_0}$.

Quite remarkably the collective activation dynamics is faithfully described in the overdamped limit and in the context of the harmonic approximation, for which the above equations simplify into:
\bseq \label{eq:Baconnier_ha}
\begin{align}
\dot\ub_i &= \Pi \hnb_i  + \sum_{j} \mathds{M}_{ij} \ub_j \\
\dot\hnb_i &= (\hnb_i \times \dot\ub_i) \times \hnb_i + \sqrt{2 D} \xi \boldsymbol{\hat{n}}_{i}^{\perp},
\end{align}
\eseq
where $\ub_i = \rb_i -\rb_i^0$ is the displacement of the agent $i$ away from its reference configuration, now expressed in units of $l_a$, $\mathds{M}$ is the dynamical matrix of the underlying elastic structure and $\Pi = \beta \tilde F_a =l_e/l_a$.
This simplification allows the authors to show that the mode selection results from the nonlinear elasto-active feedback, resulting from self-alignment, which connects the linear destabilization of the fixed points to the spatial extension and the orthogonality of the polarization of the selected modes.

Performing simulations for large system sizes, the transition from the disordered phase to the synchronized chiral oscillations is observed to be discontinuous, with a transitional regime of spatial coexistence between the two phases that is controlled by the pinning condition. The authors propose a coarse grained version of the dynamical equation:
\begin{subequations} 
\label{eq:cg}
\begin{align}
\partial_t{\boldsymbol{u}} &= \Pi \boldsymbol{m} + \boldsymbol{F}_{e} \label{eq1:cg} \\
\partial_t{\boldsymbol{m}} &= (\boldsymbol{m}\!\times \!\partial_t{\boldsymbol{u}} )\!\times \!\boldsymbol{m}\!+\! \frac{1\!-\!\boldsymbol{m}^2}{2}\partial_t{\boldsymbol{u}} \!-\!D_r \boldsymbol{m},
 \label{eq2:cg} 
\end{align}
\end{subequations}
where $\ub({\bf r},t)$ and $\mb({\bf r},t)$ are the local averages of, respectively, the microscopic displacements $\ub_i$ and polarizations $\hnb_i$, the elastic force $\Fb_e\left[\ub\right]$ is given by the choice of a constitutive relation and the relaxation term $-D_r \mb$ results from the noise. Note the second term in equation~\eqref{eq2:cg}, which arises from the coarse graining procedure, allows the displacement's rate to polarize the elastic medium. Solving these equations for homogeneous solutions, one finds two coexisting solutions, disconnected in phase space: a static disordered one with zero magnetization, and a strongly magnetized oscillating chiral one.

\subsection{Active Jamming}
In the light of the above results, one would expect similar collective actuation to also take place in jammed packings of soft active particles with self-alignment. The system being jammed, its structure is essentially frozen, and thereforte the dynamics should not differ significantly from the one reported above. Active jamming was actually proposed as a first step to describe the in vitro experiments on confluent monolayers of migratory epithelial and endothelial cells~\cite{Henkes2011}. The model considers $N$ polar soft disks of radius $a_i$, the position $\rb_i$ and orientation $\hnb_i$ of which evolve with the overdamped equations of motion:%
\bseq  
\label{eq:Henkes}
\begin{align}
	\dot\rb_i &= v_0 \hnb_i + \alpha  \sum_{j=1}^{z_i} \vec{F}_{ij},\\
	\dot{\theta}_i &= \frac{1}{\tau_a} (\psi_i-\theta_i)+ \eta_i,
\end{align}
\eseq
Soft repulsive interaction force between $i$-th and $j$-th disks is defined by $\Fb_{ij}=-k(a_i+a_j-r_{ij})\heb_{ij}$ if $r_{ij}< a_i + a_j$ and $\Fb_{ij}=0$ otherwise. The self-alignment term is identical to the one introduced in~\cite{Szabo2006} and corresponds to the linearization of the self-alignment in the case where $\wb = \hvb$.
The angular noise $\eta_i$ is considered as Gaussian with zero mean and variance $\langle\eta_i(t)\eta_j(t^{\prime})\rangle=\sigma^2\delta_{ij}\delta(t-t^{\prime})$.

\begin{figure}[t]
\includegraphics[width=0.95\columnwidth]{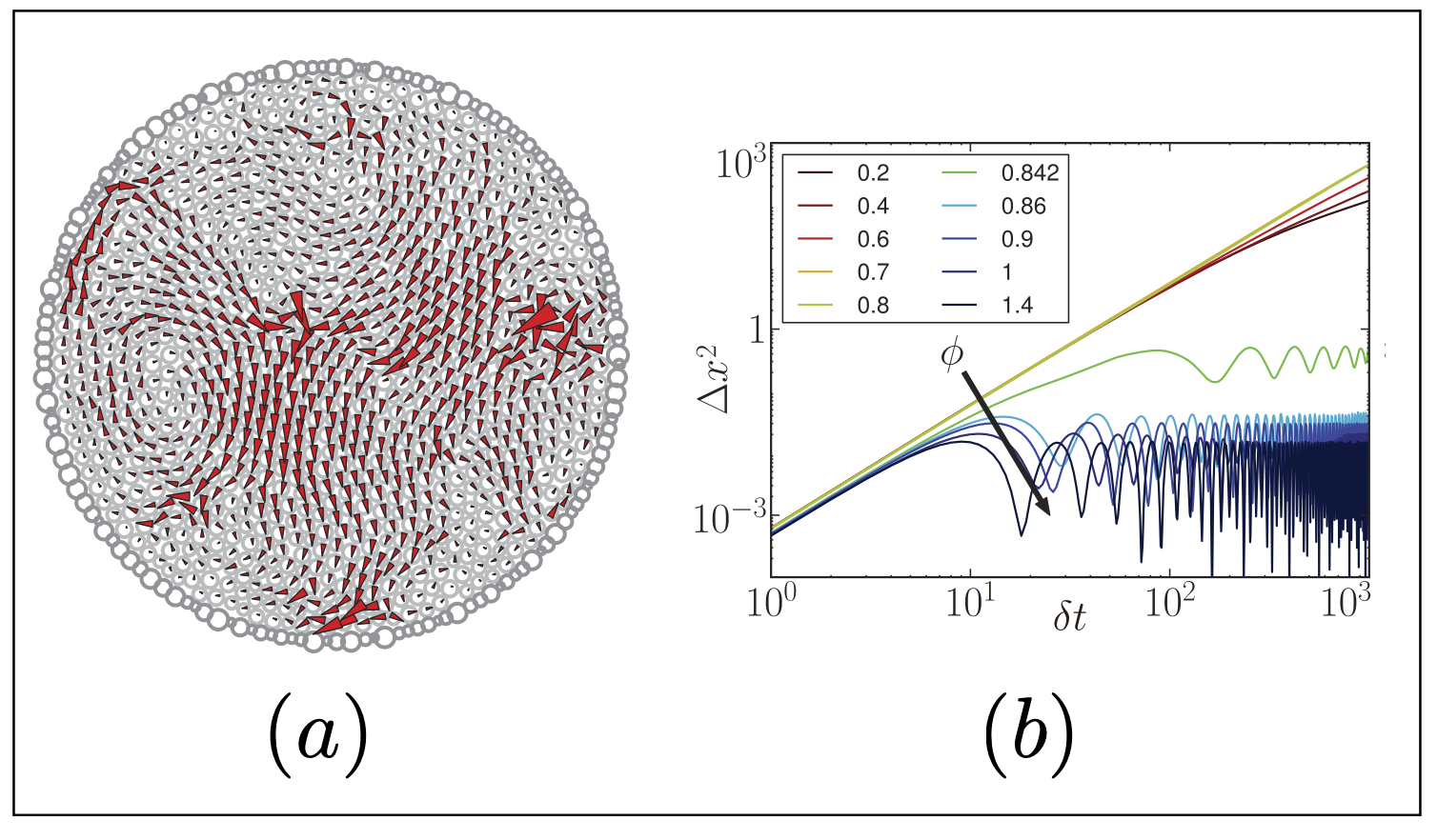}
\caption{\textbf{Collective actuation of a jammed packing of soft particles}. 
(a) A snapshot of the dynamics obtained from the simulation of a jammed packing of self-aligning soft polar disks described by equations~\ref{eq:Henkes}; the outer glued boundary is shown in dark gray; the red arrows represent the instantaneous velocity field, with $v = v_0$ corresponding to an arrow of length $1$ in units of the particle diameter; 
(a2) mean-square displacement as a function of time lag, for different packing fractions $\phi$, at fixed $v_0$, showing a transition from diffusion at low $\phi$, to polar alignment for $\phi < 0.8$ and to an oscillating jammed state around $\phi = 0.842$; adapted from~\cite{Henkes2011}.
}
\label{fig:collact-jam}
\end{figure}

This model was studied in a circular confined geometry with polydisperse soft particles to avoid unrealistic crystallisation effects in the dense phase. A self-actuating or oscillating phase emerges for small noise values at densities above the jamming transition $\phi_J \approx 0.842$ and at small self-propulsion velocities (Fig.\ref{fig:collact-jam}-a,b). At larger velocities, a uniformly rotating state emerges, followed by a re-emergence of the circular swarm state, already analyzed in~\cite{Szabo2006}, at moderate densities. 
The self-actuating state was then analyzed using a normal modes formalism by projecting the motion onto the normal modes of the packing computed at the passive, minimised potential energy state nearest to the average particle positions. The motion shows regular oscillations with a frequency that corresponds directly to the lowest available normal mode in the confined system, and with significant coupling to only 5-10 of the lowest modes. This was confirmed using a linear response calculation that maps the motion to a damped oscillator in the normal modes, with a frequency proportional to mode stiffness $K_{\nu}$ and damping that vanishes for the lowest excited mode.

%
%\begin{equation} 
%\ddot{a}_{\nu} + D_r \left[1-\frac{v_0 \Delta}{v_{\text{rms}}} + \frac{\alpha}{D_r} K_{\nu} \right ]\dot{a}_{\nu} + \alpha D_r K_{\nu} a_{\nu} = 0. 
%\end{equation}
%

\subsection{Active Voronoi models in confinement}
\label{subsec:SPV}
\begin{figure}
\centering
\includegraphics[width=0.95\columnwidth]{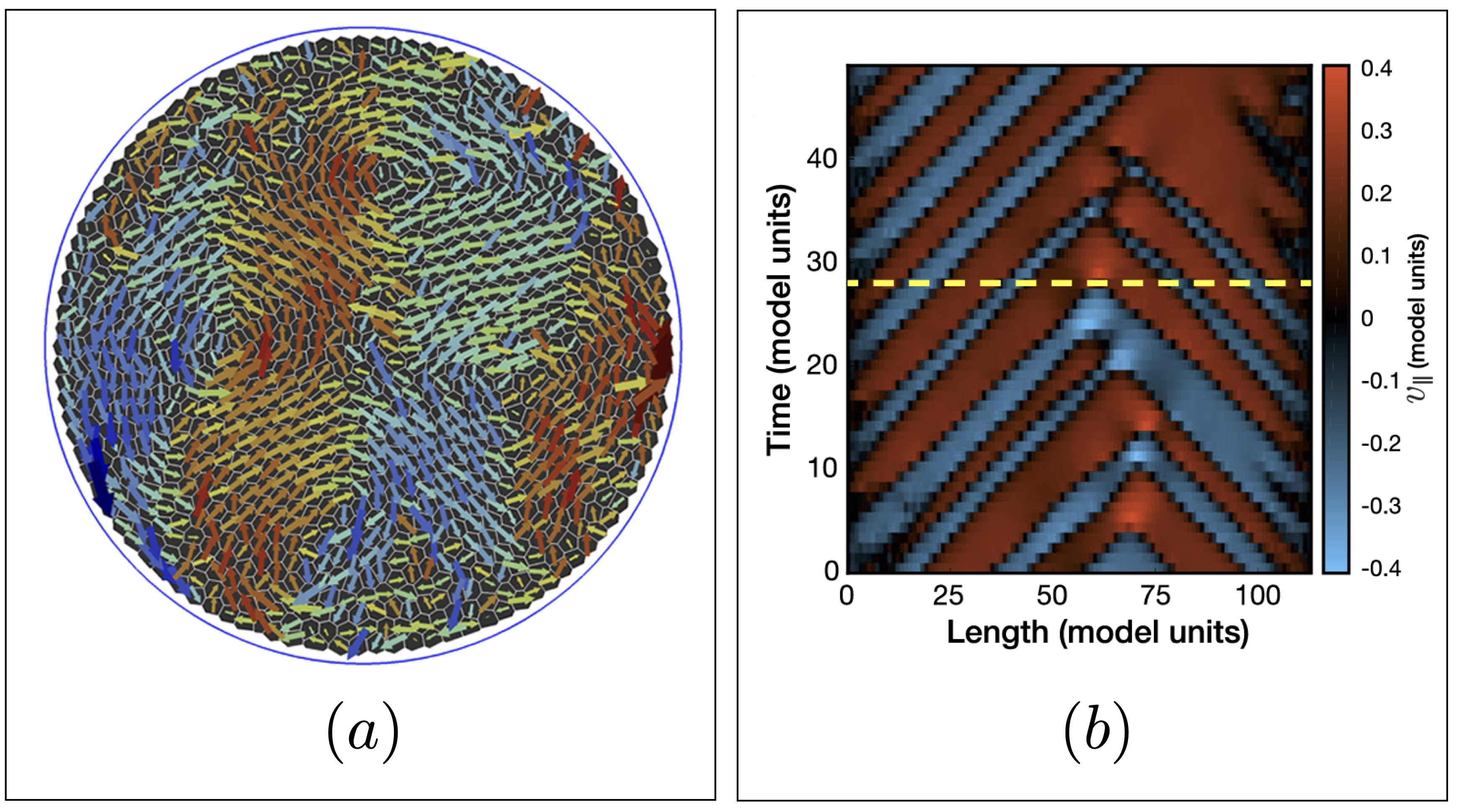}
\caption{\textbf{Collective actuation in a self-propelled Voronoi model in confinement:} (a) Simulated tissue in circular confinement in the solid flocking phase, showing the velocity field in steady-state oscillation ($p_0=3.385$ and $J=1$); (adapted from~\cite{Barton2017}); (b) Kymograph (space-time plot) of the velocity along a quasi one-dimensional channel of simulated cells; (adapted from~\cite{Petrolli2019}).}
\label{fig:collact_voronoi}
\end{figure}

The self-propelled Voronoi model as introduced in section~\ref{sec:ActiveVoronoi} and eq.~(\ref{eq:AJVoronoi}) is a more realistic model of collective cell mechanics than the simple particles of \cite{Szabo2006, Henkes2011}. A good model for tissue oscillations can thus be obtained by combining the SPV model with self-alignment and explicit tissue boundaries that allow for confinement. This was first implemented numerically in~\cite{Barton2017}. For SPV parameters that belong to a solid flocking phase in flat space, the model indeed leads to steady state oscillations in confinement with a phenomenology that is very similar to the active jamming described in~\cite{Henkes2011} (Fig.~\ref{fig:collact_voronoi}(a)). 
A significant step further was achieved in~\cite{Petrolli2019} when matching this model to experiment by simulating long, thin cell sheets with fixed boundaries to match the long, thin experimental setup with cells confined to a quasi 1d channel by selective coating with fibronectin. Similar to the experiment shown in figure~\ref{fig:collact_bio-tissues}(a), the model kymograph shown in Figure~\ref{fig:collact_voronoi}(b) shows regular oscillatory phenomenology.
%Further work is necessary to make a full, quantitative match between experiment and simulation, and it will be interesting to see if there are equivalents to the collective and global alternating rotation oscillatory states here.

\subsection{Mechanical oscillations in bacterial colonies and tissues}

\begin{figure}
\includegraphics[width=0.95\columnwidth]{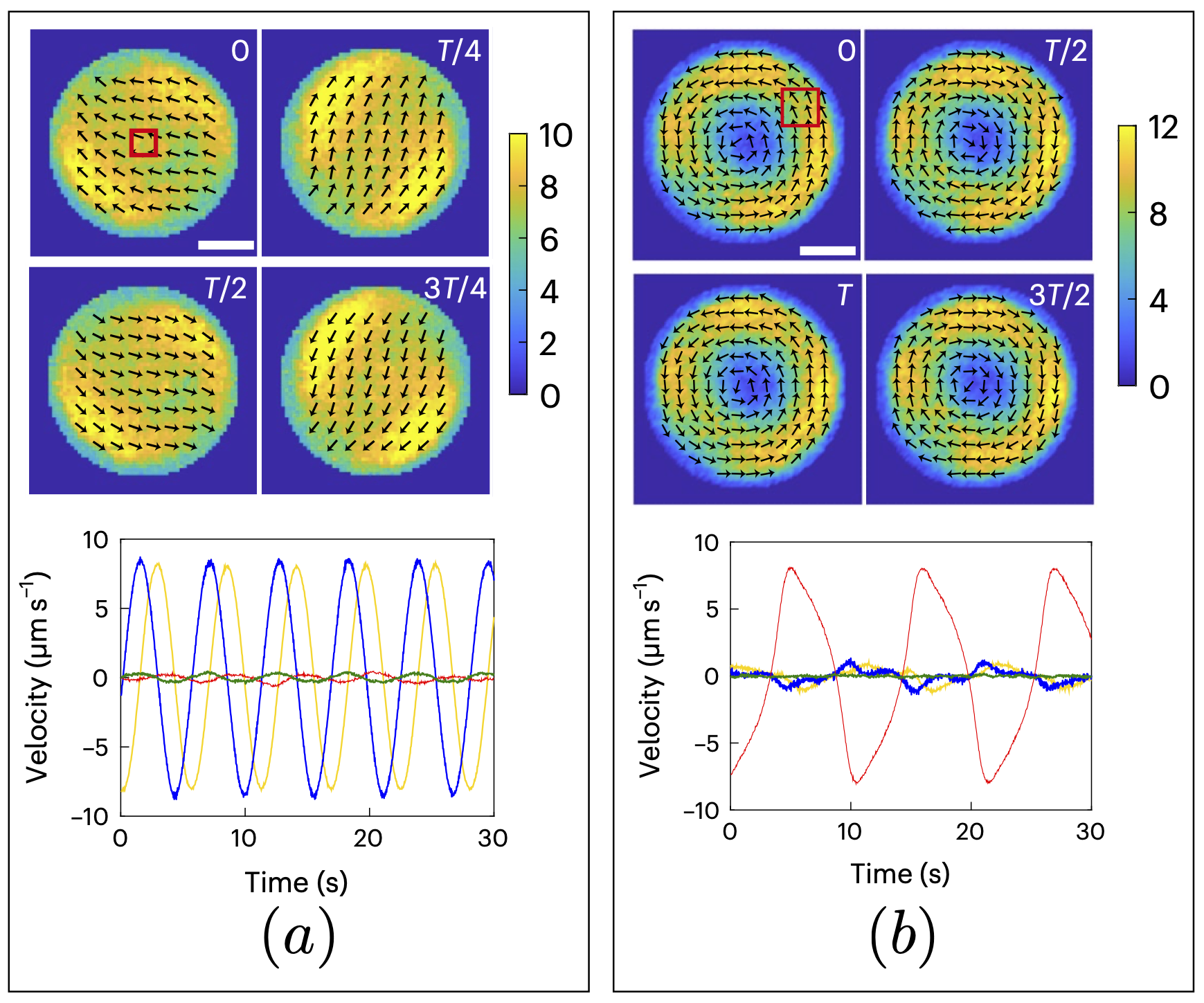}
\caption{
{\bf Collective oscillations in bacterial biofilms}. 
(a) Oscillatory translation along circular orbits and (b) oscillatory global rotation; Top: time sequences of collective velocity field. Arrows represent the velocity directions, and the colormap indicates the velocity magnitude in $\mu$ m/s; scale bar: $500 \mu$m. Bottom: temporal dynamics of the spatially averaged collective velocity decomposed into Cartesian (yellow and blue traces) and polar-coordinate (red: azimuthal, green: radial) components (adapted from~\cite{Xu2023}).
}
\label{fig:collact_bio-bacteria}
\end{figure}

Quite remarkably, the two collective actuation dynamics reported in the experimental model of active elastic structure, namely the synchronized chiral oscillations, and the global alternating rotation, were observed in a single system of millimetre-sized, quasi-2D and disk-shaped \textit{P. mirabilis} biofilms~\cite{Xu2023} (Fig.~\ref{fig:collact_bio-bacteria}-a,b). The biofilm, with the top surface exposed to air and the bottom surface in contact with agar that provides nutrients and substrate adhesion, is laterally confined by a rim of immobile bacteria. Under isotropic confinement, it exhibits the two topologically distinct global dynamics reported above, and a transition between the two regimes, when tuning the activity level. Similar control was achieved in the case of the toy model elastic structure~\cite{Baconnier2023}. The biofilm also exhibits self-sustained elastic waves with a power-law scaling of the wave speed with activity~\cite{Xu2023}. 
The authors reproduce their finding in a model of overdamped self-aligning self-propelled particles connected by Hookean springs in a 2$d$ triangular lattice, which is very close to the one introduced in~\cite{Baconnier2022} to describe the model experimental active elastic structures.
%
%\bseq  \label{eq:Xu}
%\begin{align}
%\dot{\boldsymbol{r}}_{i} &= v_0 \boldsymbol{n}_{i} + \Xi_i %\left( \boldsymbol{F}_i + D_r \boldsymbol{\hat{\xi}}_r\right), %\label{eq1:xu_model} \\
% \dot{\boldsymbol{n}}_{i} &= \beta \left[ \left( \boldsymbol{F}_i + D_r \boldsymbol{\hat{\xi}}_r\right) \cdot \boldsymbol{n}_i^{\perp} \right] \boldsymbol{n}_i^{\perp} + D_{\theta}\boldsymbol{\hat{\xi}}_{\theta} - \Gamma \frac{\delta F_n}{\delta \boldsymbol{n}_i}, \label{eq2:xu_model} 
%\end{align}
%\eseq
%
The global alternating rotation dynamics was also observed in a dense active suspension of Escherichia coli confined in a visco-elastic fluid, the rheology of which is controlled by the addition of purified genomic DNA~\cite{Liu2021}.
This behaviour is explained by the interplay between active forcing and viscoelastic stress relaxation without explicit reference to a self-alignment mechanism.
However, as the authors use a coarse grained model, this explanation does not preclude the existence of self-alignment at the microscopic scale.

\begin{figure}
\includegraphics[width=0.95\columnwidth]{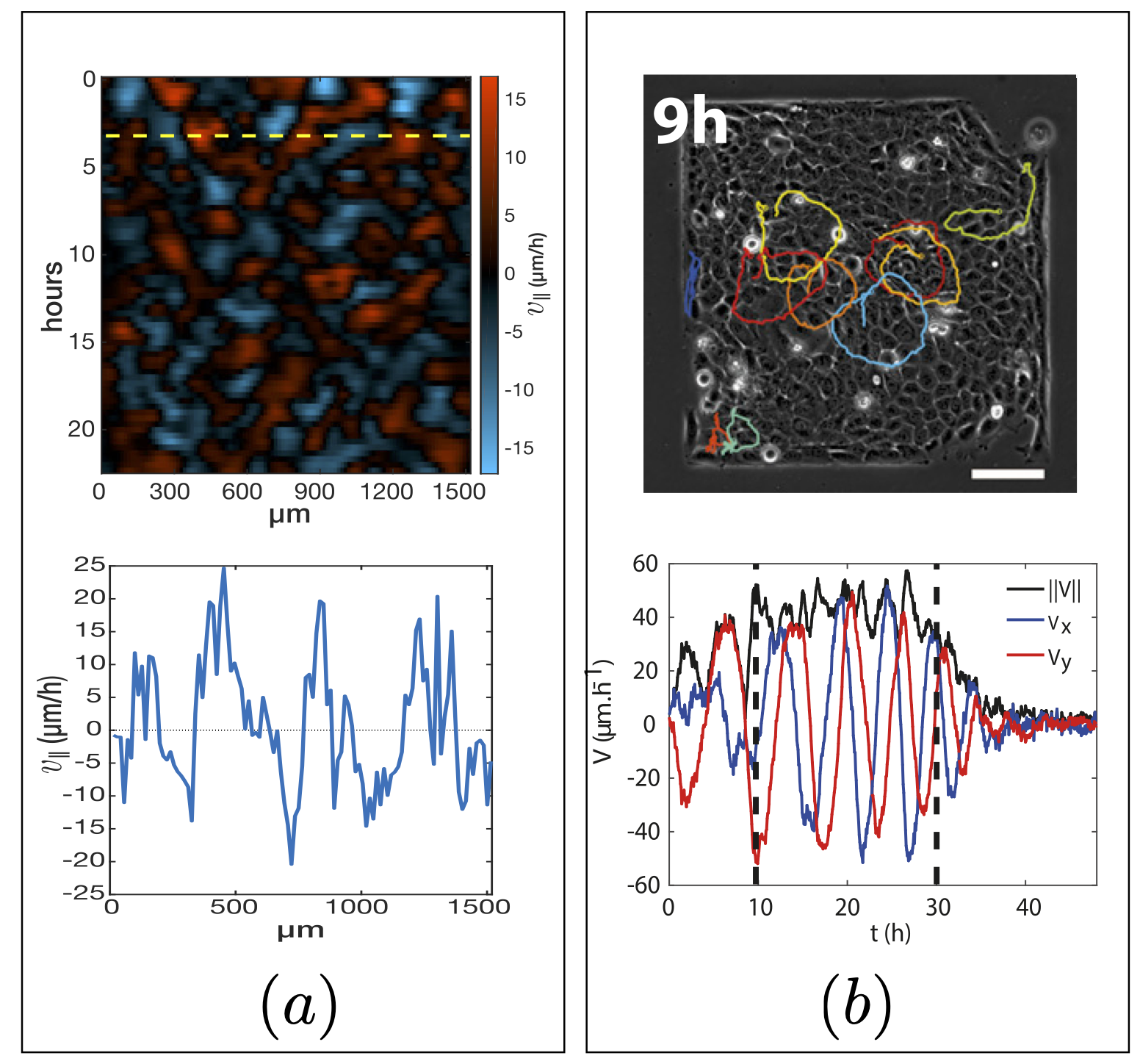}
\caption{
{\bf Collective oscillations in tissues}. (a) The velocity field of MDCK cells seeded onto a polyacrylamide (PA) gel patterned with fibronectin stripes are measured by PIV; Top: A kymograph representing the average horizontal velocity $v_{\parallel}(x; t)$ demonstrates the presence of spatial and horizontal oscillations; Bottom: an example of the velocity profile along the dashed line (adapted from~\cite{Petrolli2019};
(b) Dynamics of a confluent HaCaT monolayer in square confinement; Top: snapshots of the cell monolayer superimposed with representative trajectories of single cells performing translational oscillations along circular orbits; scale bar: $100$ mm. Bottom time evolution of the two projected components, $V_x = \langle v_x \rangle_{\text{ROI}}$ and $V_y = \langle v_y \rangle_{\text{ROI}}$, and norm of the velocity $||V|| = \left( V_x^2 + V_y^2 \right)^{1/2}$, computed on a cropped area in the center of the square of the Top panel, (adapted from~\cite{Peyret2019}).
}
\label{fig:collact_bio-tissues}
\end{figure}

Epithelial cell sheets are often subjected to large-scale deformation during tissue formation and the active mechanical environment in which they operate is known to promote collective oscillations. For instance epithelial monolayers of Madin-Darby canine kidney (MDCK) cells in circular confinement exhibit swirling flows, which rely on the interplay of confinement and local alignment~\cite{Doxzen2013}. This swirling flow can be supplemented by a sub-dominant radial oscillation~\cite{Deforet2014,Notbohm2016}. 
Analyzing the collective motion of epithelial cells confined to a quasi-one-dimensional channel in the light of numerical simulations, based on a self-propelled Voronoi model (see section~\ref{subsec:SPV}), recent works~\cite{Petrolli2019} show that tuning the length of the confining channel drives a phase transition between a state of global oscillations and a multinodal wave state (Fig:~\ref{fig:collact_bio-tissues}-a). This transition is a consequence of the interplay between local cell active dynamics and global confinement. The effect is demonstrated to rely on self-alignement.
Also spectacular are the dynamics reported in the case of human keratinocytes (HaCaT) and enterocytes (Caco2), where all cells perform a synchronized chiral oscillation~\cite{Peyret2019} (Fig:~\ref{fig:collact_bio-tissues}-b). Using molecular perturbations, the authors demonstrate that force transmission at cell-cell junctions and its coupling to cell polarity are pivotal for the generation of these collective dynamics. This system was then modelled using self-alignment dynamics with a phase-field model of the tissue, where each phase corresponds to an individual deformable cell.

\section{Discussion and Perspectives}
\label{sec:discussion}

\subsection{A unified framework ?}

The models discussed in the present review (see Table~\ref{table:all_models}) distinguish themselves by the way elasticity emerges, and by the type of boundary condition. But most importantly, they can be distinguished according to whether 
\begin{itemize}
    \item the alignment strength is proportional to the amplitude of the velocity, $\wb = \vb = \dot\rb$, or not, $\wb = \hvb$,
	\item the aligning torque is linearized or not (L / NL),
 	\item the damping and the translational noise are taken isotropic or not (Y / N).
\end{itemize}

The experiment described in section~\ref{subsec:walkers}, where a hexbug is driven along the side of a square, demonstrated that it should be described with an alignment torque proportional to the amplitude of the velocity. When describing the dynamics  of a hexbug in an harmonic potential~\cite{Dauchot2019} adopted the same choice. A simple model of a mechanical walker, described as a rigid active body with a distribution of friction forces $\gamma(\boldsymbol{r})$ off-centered with respect to its center of mass leads to a similar self-alignment when the body is symmetric with respect to $\hnb$.

Generally, for a mechanical walker characterized by a specific self-propulsion density $\boldsymbol{a}(\boldsymbol{r})$ and subjected to a set of external forces with a unique center of rotation  $\boldsymbol{r}_{cr}$ over time — defined as the point where the total external torque is nullified — the overdamped equations of motion can be reformulated in a simplified manner (see details in Appendix~\ref{section:appendixA}). When the rigid walker possesses an axis of symmetry, its dynamics are determined by:
\bseq 
\begin{align} 
 \label{centerConti_main}
 {\boldsymbol{v}}&=v_0\hat{\boldsymbol{n}}+\frac{1}{\gamma}\Fbext,\\
\label{angularConti_main}
    \dot{\theta}&= \frac{\gamma (\boldsymbol{r}_{cr}-\boldsymbol{r}_{cf})\cdot\hat{\boldsymbol{n}}}{I_\gamma } \left[    \hat{\boldsymbol{n}}\times\boldsymbol{v} \right]\cdot\hat{z}.
\end{align}
\eseq
The velocity is given in terms of the center of friction $\boldsymbol{v}={\dot{\boldsymbol{r}}_{cf}}$, where $\boldsymbol{r}_{cf}=\frac{1}{ \gamma}\int_\Omega  \rb \gamma(\rb) d\rb$ and $\gamma=\int_\Omega \gamma(\rb) d\rb$ is the effective friction coefficient.
The net self-propulsion is $v_0\hat{\boldsymbol{n}}=\frac{1}{ \gamma}\int_\Omega  \boldsymbol{a}(\rb) d\rb$, with $\hat{\boldsymbol{n}}$ being the unit vector defining the orientation of the walker, and $I_\gamma= \int_\Omega  \gamma(\rb) (\rb-\rb_{cf})^2 d\rb$ is the moment of friction measured from the center of friction. The total external force $\Fbext$ could originate from gravity to contact interaction forces when agents are frictionless disks; in the first case, the center of rotation is given by the center of mass, while in the latter, it is at the center of the disk.

A key requirement to induce self-alignment in the model (Eq.~\eqref{angularConti_main}) is that the center of rotation  $\boldsymbol{r}_{cr}$ differs from the center of friction $\boldsymbol{r}_{cf}$. Remarkably, the aligner or anti-aligner nature arises from the parallel or anti-parallel character of the self-propulsion direction $\hat{\boldsymbol{n}}$ with respect to the vector $(\boldsymbol{r}_{cr}-\boldsymbol{r}_{cf})$. This simple model adds robustness to the nonlinear model with $\wb = \vb$. 

In the supplementary material of ~\cite{Dauchot2019}  it was shown that making the alternative choice $\wb=\hvb$ suppresses the frozen, climbing, state, of a hexbug in an harmonic potential and the orbiting state is observed for any positive value of the coupling parameter. However the minimal ingredients of a microscopic model which leads to such different macroscopic self-alignments remain to be explored.

To assess the generality of these results, we introduce a more general model for the angular dynamics:
\begin{equation}
    \dot{\theta} = \beta |\vb|^\nu f\left[\left(\hnb \times \hat\vb\right)\cdot \hat{z}\right].
\end{equation}
The choice $\nu=1$ corresponds to $\wb=\vb$, while $\nu=0$ corresponds to $\wb=\hvb$.
The function $f[x]$ is given by $f[x]=x$ when the aligning torque is not linearized (NL), and by $f[x]=\sin^{-1}(x)$ when it is (L); in general it should be a non-singular odd function with a domain given by $[-1,1]$.
Analyzing the stationary states and the stability of the frozen state of a self-propelled particle with such angular dynamics in an harmonic potential (appendix~\ref{app:harmonic_potential}), we find that the behavior is largely independent of the particular choice of $f[x]$: for $\nu<1$ chiral states are observed for all the values of the alignment parameter $\beta$, while for $\nu=1$ they are observed only beyond a threshold $\beta_c$.
For $\nu>1$, multiple chiral states may exist.

\begin{figure}[t!]
\includegraphics[width=0.95\columnwidth]{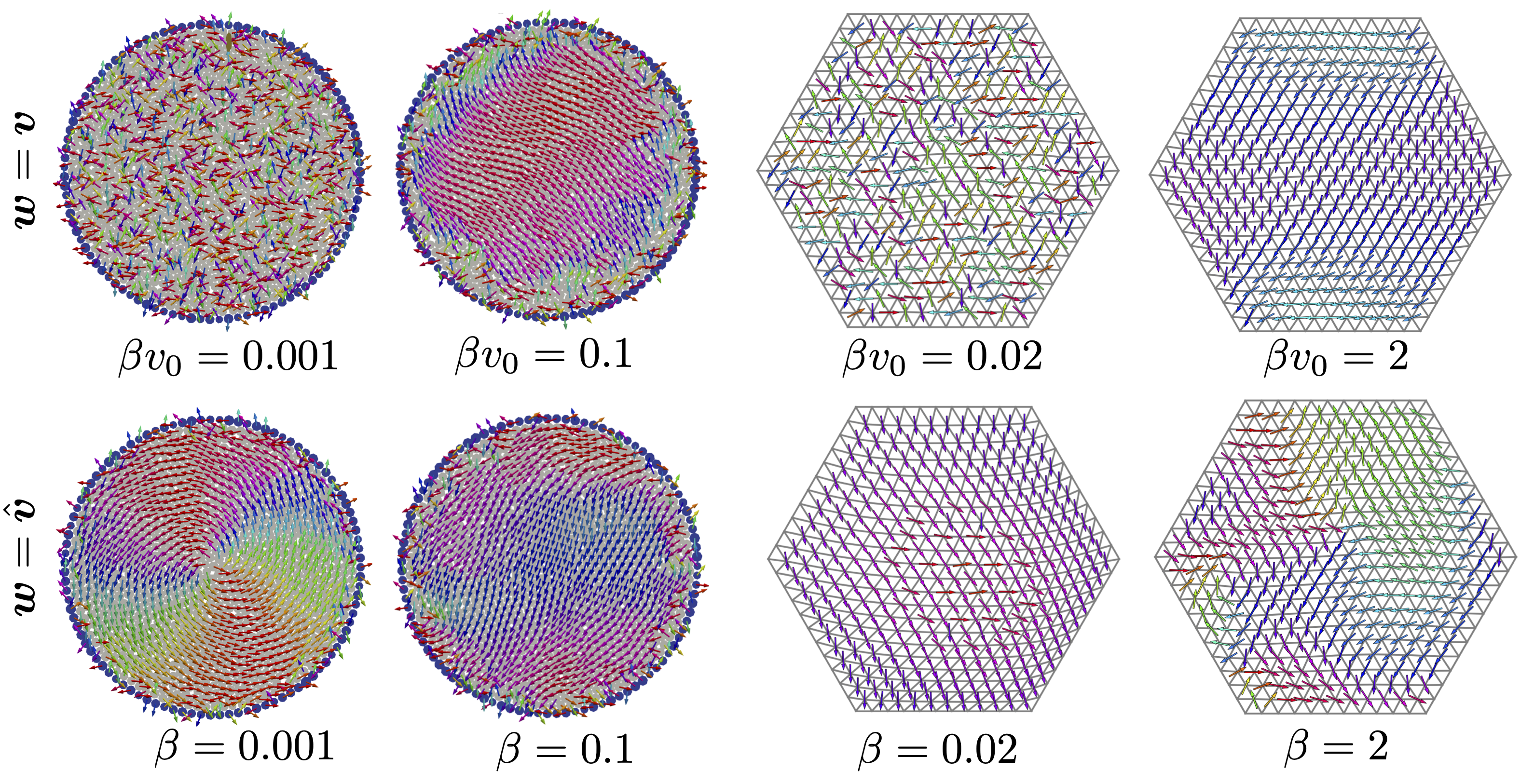}
\caption{{\bf Collective actuation in particle systems (left) and elastic network (right) for $\wb = \vb$ and $\wb = \hvb$, respectively}. Obtained from simulating equations~\ref{eq:New_od} with the appropriate interactions (harmonic repulsion or harmonic springs) and in the limit of very small angular noise.}
\label{fig:sa-v-vs-vhat}
\end{figure}

At the collective level, synchronous oscillations were reported whether $\wb = \vb$ or $\wb = \hvb$, however for different systems. The case $\wb = \vb $ was discussed in the context of elastic structures, while that with $\wb = \hvb$ was considered for dense assemblies of soft particles. To illustrate this point, we have run simulations of equations~\eqref{eq:New_od}, with $\wb = \vb$ and $\wb = \hvb$, in jammed assemblies of particles and elastic network (Fig: \ref{fig:sa-v-vs-vhat}). Independently of the system of interest, when $\wb = \vb$, there is a regime at small self-alignment where collective actuation does not take place and the transition first reported in the case of elastic structures, is also observed in the case of jammed assemblies of soft particles, while in the case of $\wb = \hvb$ collective actuation takes place for arbitrary small values of the self-alignment. Hence the collective behavior essentially inherits the property analyzed at the single particle level.

\subsection{Large scale hydrodynamics theory}

Large scale hydrodynamics formulations of dry active matter have been introduced as early as 1995 to describe the flocking phase of polar liquids~\cite{Toner1995}. Soon after, the case of active nematic liquids on a substrate was also considered~\cite{Ramaswamy2003}. In both cases, the phases that are considered are liquids of particles taking their momentum from the ground. 
The theory of active gels, on the contrary, was introduced to describe the mechanical behavior of cells, assuming momentum conservation~\cite{Kruse2004,Kruse2005}. Also, it accounts for a short time elastic regime, that is captured by introducing a Maxwell timescale in the constitutive relation.
All such hydrodynamic theories can be obtained from a perturbative expansion around a specific ordered state, the existence of which is not guaranteed. An alternative strategy is to develop the theory in the vicinity of an equilibrium state, at the cost of leaving aside systems that are truly far from equilibrium. Finally, one can write down a microscopic theory and coarse grain it to obtain the hydrodynamic equations for the large scale fields. This is usually a difficult task paved with approximations that are not always well controlled. Nevertheless active hydrodynamic theories have the advantage of being generic in the sense that they describe different systems provided that they share the same conservation laws and symmetries. 

The polar systems discussed in the context of the present review take their momentum from the ground and therefore belong to the so-called class of dry polar active matter, for which there is no conservation of momentum. Their liquid phase is therefore described by Toner-Tu like equations~\cite{Toner1995,Toner2005} and the Landau terms of these equations were explicitly derived for self-aligning disks, demonstrating that self-alignment effectively leads to collective motion~\cite{Lam2015,Lam2015a}, as discussed in section~\ref{sec:sa_liquids}. 

The case of active solids was considered only more recently. A first approach is to couple phase-field crystals with the Toner-Tu equations to describe solid polar flocks~\cite{Menzel2013,Alaimo2016,Ophaus2018}. This however ignores a symmetry-mandated coupling between orientation and strain, due to rotation invariance. In passive orientationally ordered solids, such a coupling leads to the vanishing of the zero-frequency shear modulus~\cite{Dalhaimer2007}. The role of activity includes the possibility of stable uniaxial active solids when momentum is not conserved, and the instability of the uniaxially ordered phase in momentum-conserved systems for large active forcing~\cite{Maitra2019}. In all cases, such theories of active solids have so far considered oriented active solids, that is active solids with polar or nematic order. In the context of the present review, they predict the stability at large scale of collective motion in the active sheet.

A complementary approach is to considered scalar active solids, where activity stems from non conservative interaction forces~\cite{Scheibner2020}. Considering the case where momentum is conserved, the forces derive from a stress tensor, which in the limit of small strain is linearly related to the strain tensor by the static elastic modulus tensor, as it is the case for usual passive elastic materials. However static elastic moduli altogether absent in passive elasticity can arise from the active, non-conservative microscopic interactions. These active moduli enter the antisymmetric part of the static elastic modulus tensor, hence the name ``odd elasticity'' and quantify the amount of work extracted along quasistatic strain cycles. The application of these ideas to the systems described here is not straightforward because the absence of momentum conservation gives rise to forces which do not derive from a stress tensor. However the self-alignment is likely to contribute to the active part of the stress tensor including the odd part of it.

Altogether the large scale physics of active systems composed of self-aligning active agents remains largely unexplored. In the liquid polar state, the computation of the Landau terms of the Toner-Tu equation demonstrates that the mean field transition to collective motion can be first order in sharp contrast with the case of the Vicsek model. The consequences of it beyond mean field remain unknown. The solid polar state is stable at large scale~\cite{Maitra2019}; however the role of structural defects remains unexplored. The mechanical response of such active solids, with the possibility of odd-elasticity, also remains unexplored. Finally the fate of the transition to collective actuation at large scale is unknown. One possible avenue for making progress in this last direction is the possible connection with non reciprocal phase transitions.

\subsection{Connection with non-reciprocal phase transitions}
\label{subsection:non_reciprocal}

It was recently shown that systems composed of microscopic degrees of freedom experiencing non-symmetrical interactions, together with non-conservative dynamics, are prone to develop chiral phases via a specific kind of phase transition called non-reciprocal~\cite{Fruchart2021}, controlled by the presence of exceptional points. 

Self-alignment can be seen as a non-reciprocal internal coupling between the displacement and the orientation within each active particle: the polarity aligns with the displacement rate, while the displacement evolves following the sum of active and elastic forces. 
At the level of a single active particle in a harmonic potential, in the case $\wb = \vb$, the polarity and displacement vectors explicitly experience non-symmetrical interactions: eliminating adiabatically the dynamics of the amplitude of the displacement, one finds that the phase of the displacement vector is chasing that of the polarity~\cite{Henkes2011,Baconnier2023}. It is also found that the transition not only takes place through an exceptional point, but that the whole orbiting, or chiral, phase stands on a line of exceptional points in the parameter space. 
At the level of the coarse-grained equations~\eqref{eq:cg}, linearizing them around the disordered phase and mapping them to the most general equations one can write for rotationally symmetric vectorial order parameters, one finds that the macroscopic displacement and polarity fields also couple non-symmetrically~\cite{Baconnier2022}. 

Non reciprocal phase transitions could therefore be a good paradigm to discuss the transition to collective actuation. A more precise analysis however remains to be done. In particular the transition considered here takes place from the disordered to the chiral phase, while that discussed in the context of non reciprocal phase transitions arises from the polar or anti-polar phase.

\subsection{Further Perspectives}
We conclude this review by offering a few perspectives, which we believe are promising research avenues.

At the conceptual level, the increasing importance of self-alignment as a generic mechanism for collective actuation in active solids is likely to raise many theoretical developments. As mentioned above, a general continuous description of dry active solids is still missing, not to mention the mechanical properties of such solids, from linear elasticity to plasticity and failure. Remarkably, the stability of collective actuation in the thermodynamic limit remains to be investigated. Since collective actuation results from the coupling of a spin like degree of freedom to the vibration of the elastic structure, it could be interesting to probe the response to an external polarizing field: controlling the orientation of the agents, with an external field could be a practical way of controlling the actuation locally.

In a different vein, it has recently been shown that bulk or boundary disorder are very relevant to active systems because of the presence of long-ranged density and current modulations~\cite{Morin2017,Besse2022,Granek2023,Benvegnen2023}. This is also true for any amount of anisotropy, which drastically alters the phenomenology of flocking from that of the rotationally invariant case~\cite{Solon2022}.
The precise properties of collective actuation in the presence of disorder are a central issue for any practical application. Different types of disorder must be considered: The elastic structure can for instance include several types of defects, or instead correspond to a inherently disordered jammed or glassy packing. One may also want to consider disorder in the self-aligning strength, or the effect of a random external field. Given that elastic modes will couple differently in the presence of disorder, one can expect new fascinating behaviors, such as localization of the actuation.

A last general question concerns the sign of the self-aligning torque. It was realised only very recently that negative self-alignment can lead to interesting dynamics~\cite{BenZion2023} and there has yet been no systematic study of the collective dynamics of a large assembly of self-anti-aligning particles. When two such particles collide, they point toward each other leading to an effective adhesion that can be much stronger than in the case of scalar active matter. It is thus likely that such particles will experience Motility Induced Phase Separation at very low packing fraction, even for moderate persistence time. Also, to date the coupling of anti-alignment with elastic interactions remains completely unexplored.

In the context of swarm robotics, it was shown that the sign of self-alignment can be influential in the way a swarm of robots solves a simple phototactic task, with strong steric frustration~\cite{BenZion2023}, and make it an essential ingredient of self-organization and function. For example, in biological organisms like \textit{Trichoplax Adhaerens}, self-alignment is thought to play a role in setting large-scale coordination and complex forms of locomotion \cite{davidescu2023growth}. The complex dynamics and the different modes of collective motion of elastic structures doped with self-aligning active agents are thus an interesting direction to create autonomous robots exploring their environment and interacting with it.

In dense biological systems, like cell monolayers and bacterial biofilms, self-alignment is not the only contribution to the repolarisation of the microscopic agents. For monolayers, explicit active coupling between contractile junctions with nematic symmetry and mechanochemical feedback need to be considered \cite{sknepnek2023generating,rozman2023shape}, and in bacterial biofilms explicit nematic shape alignment has to be taken into account. It begs the question of how informative the empirically observed emerging collective dynamics on the properties of the microscopic constituents is. Can one state that self-alignment is present or dominant, even effectively, if observing collective actuation?   Thus constructing an explicit test for self-alignment becomes important. Since taking a single cell out of its environment irrevocably changes its phenotype, unlike for mechanical particles, this will need to take place at the collective level. A good avenue for this test will be to exploit the intimate relation of collective actuation to elastic normal modes and vary system size and boundary conditions to make predictions. 

\section{Conclusion}

In this review, we have reported, as exhaustively as we could, the numerous occasions where self-alignment manifests itself in the field of polar active matter. Although this generic, mechanically rooted, coupling between the polarity and the velocity of a self-propelled particle was introduced as early as 1996, it remained overlooked for a long time, mostly because it was thought that its specificity could be integrated out into some form of mutual alignment at the effective level. This vision, which is partly correct when considering the emergence of collective motion, misses the potentiality offered by self-alignment when considering dense assemblies of self-propelled particles or elastic networks. In that context, where the particle is essentially confined in a local harmonic potential, the natural tendency of a self-aligning particle to perform orbiting motion, as a result of a spontaneously broken chiral symmetry, leads to truly unanticipated forms of collective dynamics. 

These new dynamics are prominent in the realm of biological systems such as bacterial raft and tissues and are likely to play a crucial role in large scale synchronization phenomena. In the context of meta-materials, the selectivity of the transition to collective actuation opens the path towards the design of specific actuation patterns and new autonomous functions. Finally, in the context of swarm robotics, self-alignment allows for simpler computation rules amongst agents, which do not need to evaluate the velocity of their neighbours.

Altogether, we firmly believe that self-alignment is opening new promising routes of research and hope that the present review will serve as a useful guide into the diversity of models that were introduced independently in the past 30 years.

\section{Appendix}
\subsection{Mechanisms for self-alignment}

\subsubsection{Alignment properties of overdamped rigid walkers} \label{section:appendixA}

We shall consider an arbitrary rigid body undergoing overdamped dynamics with a distribution of active forces acting upon it. For simplicity, the rigid body is described by a discrete system of $N$ particles embedded in a rigid network. The equations of motion for the $N$ particles are given by 
\begin{equation}
\label{motion_i}
\gamma_i \dot{\rb}_i=\boldsymbol{a}_i+\boldsymbol{F}_i^{\text{int}}+\boldsymbol{F}_i^{ext},
\end{equation}
where $i\in \{1,\ldots,N\}$. Internal force $\boldsymbol{F}_i^{\text{int}}$ are given by pairwise interactions between the particles, $\gamma_i$ and $\boldsymbol{a}_i$ are the drag coefficient and self-propulsion force for particle $i$,  respectively.
Note that each self-propulsion vector $\boldsymbol{a}_i$ has a fixed orientation with respect to the agent. 
Choosing the origin of a moving frame of reference co-rotating with the rigid solid at the center of friction, $\boldsymbol{r}_{cf}=\sum_i \gamma_i \rb_i / \sum_i \gamma_i$, the particles positions in the co-rotating frame are given by $\boldsymbol{\rho}_i=\rb_i-\boldsymbol{r}_{cf}$ and their velocity $\dot{\rb}_i=\dot{\boldsymbol{r}}_{cf}+\boldsymbol{\Omega}\times\boldsymbol{\rho}_i$, where $\boldsymbol{\Omega}$ is the angular velocity of the rigid solid.
The evolution of the center of friction is obtained by summing up equations \ref{motion_i}
\begin{equation}
\label{center_eq}
 \dot{\boldsymbol{r}}_{cf}=\frac{1}{\bar{\gamma} N}\sum_i \left( \boldsymbol{a}_i+\boldsymbol{F}_i^{ext}   \right),
\end{equation}
where the internal forces cancel out and we define $\bar{\gamma}=\frac{1}{N}\sum_i\gamma_i$. Taking the origin at the center of friction, the torque balance equation reads
\begin{equation}
\label{rotation}
\sum_i \boldsymbol{\rho}_i\times \gamma_i \dot{\rb}_i = 
\sum_i \boldsymbol{\rho}_i\times \left(\boldsymbol{a}_i+\boldsymbol{F}_i^{\text{int}}+\boldsymbol{F}_i^{ext}\right).
\end{equation}
Some simple algebra then leads to 
\begin{equation}
\sum_i\gamma_i \boldsymbol{\rho}_i\times  (\boldsymbol{\Omega}\times\boldsymbol{\rho}_i)=\sum_i \boldsymbol{\rho}_i\times \left(\boldsymbol{a}_i+\boldsymbol{F}_i^{ext}\right),\end{equation}
where we use that $\sum_i \gamma_i\boldsymbol{\rho}_i=0$ by construction and $\sum_i \boldsymbol{\rho}_i\times \boldsymbol{F}_i^{\text{int}}=0$ for pairwise interaction forces. 

Finally, considering the case of a two-dimensional solid $\boldsymbol{\Omega}=\dot{\theta}\hat{z}$ and previous equations reads
\begin{equation}
\label{angular_eq}
    \dot{\theta}=\frac{1}{\bar{I}_\gamma }\left[\frac{1}{N}\sum_i \boldsymbol{\rho}_i\times\left(\boldsymbol{a}_i+\boldsymbol{F}_i^{ext}\right)\right]\cdot\hat{z}
\end{equation}
where $$\bar{I}_\gamma =\frac{1}{N }\sum_i \gamma_i \rho_i^2.$$

The general dynamics of the overdamped rigid solid are then given by equations (\ref{center_eq}) and (\ref{angular_eq}). The extension to a continuum rigid solid of volume $\Omega$ is direct. Considering that the solid has a friction density $\gamma(\boldsymbol{r})$ the center of friction is given by 
$$\boldsymbol{r}_{cf}=\frac{1}{ \gamma}\int_\Omega  \rb \gamma(\rb) d\rb$$
where the $\gamma=\int_\Omega  \gamma(\rb) d\rb$ is the effective friction coefficient. The center of friction defines the origin of the co-rotating frame of reference with coordinates $\boldsymbol{\rho}$. Defining the density of self-propulsion $\boldsymbol{a}(\boldsymbol{\rho})$ and the density of external force $\boldsymbol{f}^{ext}(\boldsymbol{r})$ the equations of motions for the center of frictions (\ref{center_eq}) is given by
\bseq 
\begin{align} \label{centerConti_eq2}
 \dot{\boldsymbol{r}}_{cf}&=v_0\hat{\boldsymbol{n}}+\frac{1}{\gamma}\boldsymbol{F}^{ext},
\end{align}
\eseq
where $v_0$ and $\hat{\boldsymbol{n}}$ are the norm and the unitary direction of the vector $\frac{1}{\gamma }\int_\Omega  \boldsymbol{a}(\boldsymbol{\rho})d\boldsymbol{\rho}$, respectively, and $\boldsymbol{F}^{ext}=\int_\Omega \boldsymbol{f}^{ext}(\boldsymbol{r})d\boldsymbol{\rho}$. The angular equations (\ref{angular_eq}) reads
\bseq 
\begin{align} 
\label{angularConti_eq2}
    \dot{\theta}&=\frac{1}{I_\gamma }\left[\int_\Omega  \boldsymbol{\rho}\times\left(\boldsymbol{a}(\boldsymbol{\rho})+\boldsymbol{f}^{ext}(\boldsymbol{r})\right)d\boldsymbol{\rho} \right]\cdot\hat{z},
\end{align}
\eseq
where the moment of friction $I_\gamma= \int_\Omega  \gamma(\boldsymbol{\rho}) \boldsymbol{\rho}^2 d\boldsymbol{\rho}$. 

We are interested in external forces that have a constant center of rotation in a co-rotating frame of reference i.e. the total external torque vanishes at a particular point at all times. Using the center of friction as the origin of the reference system, we will consider that the center of rotation is located at $\boldsymbol{\rho}_{cr}$, such that, $\int_\Omega  \boldsymbol{\rho}\times\boldsymbol{f}^{ext}(\boldsymbol{r})d\boldsymbol{\rho}= \boldsymbol{\rho}_{cr}\times \boldsymbol{F}^{ext}$. Among others, gravitational force $\boldsymbol{f}^{ext}(\boldsymbol{\rho})=\rho(\boldsymbol{\rho})\boldsymbol{g} $ or active agent under central external forces, as is the case for contact interaction for frictionless disk particles, belong to this class. In the first case, the center of rotation is given by the center of mass $\boldsymbol{\rho}_{cr}=\frac{1}{M}\int_\Omega  \boldsymbol{\rho}\rho(\boldsymbol{\rho})  d\boldsymbol{\rho}$ with $M=\int_\Omega  \rho(\boldsymbol{\rho})  d\boldsymbol{\rho}$, and in the second case, the center of rotation correspond to the center of the disk. For this class of external forces, the angular equations (\ref{angularConti_eq2}) can be  rewritten as 
\bseq 
\begin{align} 
\label{angularConti_eq3}
    \dot{\theta}&=\left[\boldsymbol{\omega}+  \frac{\gamma}{I_\gamma }\boldsymbol{\rho}_{cr}\times\dot{\boldsymbol{r}}_{cf} \right]\cdot\hat{z},
\end{align}
\eseq
where $\boldsymbol{\omega}=\frac{1}{I_\gamma }\int_\Omega  (\boldsymbol{\rho}-\boldsymbol{\rho}_{cr})\times\boldsymbol{a}(\boldsymbol{\rho})d\boldsymbol{\rho}$ is a constant angular velocity induced by the self-propulsion. 

%A natural extension to agent acting under central forces, as is the case for contact interaction for disk particles, display the same aligning behavior. Considering $$\boldsymbol{f}^{ext}(\boldsymbol{\rho})=\sum_i \delta(\rho-\rho_i)f_i \hat{\boldsymbol{r}}_i$$ with $\hat{\boldsymbol{r}}_i$ a radial unitary vector pointing toward a center placed at $\boldsymbol{\rho}_{cm}$ such $\boldsymbol{\rho}_i=\boldsymbol{\rho}_{cm}+r_i\hat{\boldsymbol{r}}_i$. Now $\boldsymbol{\rho}_{cm}$ is not necessarily the center of mass, for a disk is represent the vector from the center of friction toward the center of the disk. The only contribution to the torque in the angular equation (\ref{angularConti_eq2}) then comes from the total external force applied at $\boldsymbol{\rho}_{cc}$ and equations (\ref{angularConti_eq3}) and (\ref{centerConti_eq2}) remain the same. 

Most active agents are considered with a symmetry axis such that the friction density $\gamma(\boldsymbol{\rho})$ and the self-propulsion density $\boldsymbol{a}(\boldsymbol{\rho})$  are symmetric under reflection and the center of rotation $\boldsymbol{\rho}_{cr}$ lies along the axis of symmetry. Under this mirror symmetry, the angular velocity $\boldsymbol{\omega}$ vanishes and the center of rotation $\boldsymbol{\rho}_{cr}$ became parallel with the unit vector $\hat{\boldsymbol{n}}$, hence we recover the  form 
\bseq 
\begin{align} 
\label{angularConti_eq4}
    \dot{\theta}&= \frac{\gamma (\boldsymbol{\rho}_{cr}\cdot\hat{\boldsymbol{n}})}{I_\gamma } \left[    \hat{\boldsymbol{n}}\times\dot{\boldsymbol{r}}_{cf} \right]\cdot\hat{z}.
\end{align}
\eseq
Considering a lab frame of reference $\boldsymbol{\rho}_{cr}=\boldsymbol{r}_{cr}-\boldsymbol{r}_{cf}$.

% Tentative subsection title:
\subsubsection{Alignment properties through nonaxisymmetric forces}
\label{section:appendixA2}
Self alignment can also emerge from other asymmetries. We consider a total of $N$ off-centered disks, either rigidly connected with elastic springs (rigid network) or elastically interacting with only their repulsive counterparts. The `off-centered' aspect suggests that the center of mass of these disks is not at their geometric center. The center of mass of the $i$-th particle $\rb^{\textsc{cm}}_i$ defined as $\rb^{\textsc{cm}}_i=\rb^{\textsc{co}}_i\pm R \hat{\nb}_i$, where $\rb^{\textsc{co}}_i$ is the position of geometric center and $R$ is distance between center of mass and geometric center. The total interaction force acts on the $i$-th disk $\fb_i = \sum_{j} \fb_{ij}(\rb^{\textsc{cm}}_i, \rb^{\textsc{cm}}_j)$. 

The self-alignment torques then emerge as explicit mechanical torques $\tau_i=\bar \beta\sum_{j} \rb^{\textsc{cm}}_i \times \fb_{ij}(\rb^{\textsc{cm}}_i, \rb^{\textsc{cm}}_j)$, where $\beta$ sets the strength of the self-alignment. This can be expressed as
\bea
\tau_{i} &=& \pm \bar \beta R \hat{\nb}_i \times \sum_{j} \fb_{ij}(\rb^{\textsc{cm}}_i, \rb^{\textsc{cm}}_j).
\label{eq:self_alignment_torque_off_centered}
\eea
In this interpretation when the center of mass is positioned towards the heading direction $\hat{\nb}_i$ i.e., $\rb^{\textsc{cm}}_i=\rb^{\textsc{co}}_i + R \hat{\nb}_i$, it leads to alignment $\tau_{i}=\bar \beta R \fb_i\cdot \hat{\nb}_{i}^{\perp}$. Conversely, when the center of mass is positioned opposite to the heading direction $\hat{\nb}_i$ i.e., $\rb^{\textsc{cm}}_i=\rb^{\textsc{co}}_i - R \hat{\nb}_i$, it results in anti-alignment $\tau_{i}=-\bar\beta R \fb_i\cdot \hat{\nb}_{i}^{\perp}$. In this scenario, alignment or anti-alignment emerges purely from the dynamics of elastically interacting objects with off-centered mass.

In overdamped dynamics, the position of the $i$-th off-centered object evolves as
\bea
\dot{\rb}_i &=& v_0 \hat{\nb}_i +\mathbb{M}_i \fb_i,
\eea
where $\mathbb{M}_i$ represents the mobility matrix, defined as $\mathbb{M}_i = \mu_{\parallel} \hnb_i \hnb_i + \mu_{\perp} (\mathbb{I}-\hnb_i \hnb_i)$. We recover the evolution of the heading direction $\hnb_i$ by letting $\beta = \bar\beta R$ as follows
\bea
\dot{\hnb}_i &=& \beta (\fb_i\cdot \hnb_{i}^{\perp}) \hnb_{i}^{\perp},
\eea
which is equivalent to $\dot{\theta}_i = \beta (\hnb_{i}\times\fb_i)\cdot \hat{z}$. In this context, self-alignment emerges due to the alignment of the heading direction $\hnb_i$ towards total interacting force $\fb_i$, unlike the alignment towards velocity $v_i$ observed in the case of overdamped rigid walkers, as presented in \ref{section:appendixA}.

\subsection{Self-aligning active particle in a harmonic potential with isotropic damping} \label{app:harmonic_potential}

\subsubsection{General model}

After rescaling, the equations of motion are given by:
\bseq \label{eq:gen_overdamped}
\begin{align}
	\dot\rb &= \hnb  - \rb, \label{eq1:gen_overdamped} \\
	\dot{\theta} &= \beta |\vb|^\nu f\left[\left(\hnb \times \hat\vb\right)\cdot \hat{z}\right]  \label{eq2:gen_overdamped}
\end{align}
\eseq
Where $f[x]$ is a non-singular odd function with a domain given by $x\in[-1,1]$. The cases considered in the literature include $f[x]=x$ and $f[x]=\sin^{-1}(x)$. The exponent $\nu$ is non-negative,  in the literature the values used are $\nu=0$ or $\nu=1$. 

Using the rotational invariance, a given state is described by the distance $r$ to the origin, and the angle $\phi$ between the position and the orientation.
If $\alpha$ is the angle of the position, $\phi$ is given by $\phi=\theta-\alpha$.
Projecting eq.~\eqref{eq1:gen_overdamped} on $\rb$ and $\rb^\perp$, we get $\dot r=\cos(\phi)-r$ and $r\dot \alpha=\sin(\phi)$.
Using these two projections, we can get the norm of the velocity, $|\vb|=\sqrt{1+r^2-2r\cos(\phi)}$ and the vector product, $\left(\hnb \times \hat\vb\right)\cdot \hat{z}=-\left(\hnb \times \rb\right)\cdot \hat{z}/|\vb|=r\sin(\phi)/\sqrt{1+r^2-2r\cos(\phi)}$.
Finally, we get the equations
\bseq\label{eq:gen_overdamped_min}
\begin{align}
    \dot r & = \cos(\phi)-r, \label{eq:gen_overdamped_min_r}\\
    \dot \phi & = \beta\left[1+r^2-2r\cos(\phi)\right]^{\nu/2}f\left[\frac{r\sin(\phi)}{\sqrt{1+r^2-2r\cos(\phi)}}\right] \nonumber\\
    &\qquad -\frac{\sin(\phi)}{r}. \label{eq:gen_overdamped_min_phi}
\end{align}
\eseq

\subsubsection{Stationary solutions}

We want to consider stationary solutions, which statisfy $\dot r=0$ and $\dot\phi=0$.
These contain the frozen state, $r=1$, $\phi=0$ and chiral states where $\phi\neq 0$.

From $\dot r=0$ we get $r=\cos(\phi)$, and inserting this relation in $\dot\phi=0$, we get
\begin{equation}
    \beta|\sin(\phi)|^{\nu/2} f \left[\frac{r\sin(\phi)}{|\sin(\phi)|}\right] = \tan(\phi). 
    \label{eq:gen_phi_stat}
\end{equation}
Since $f(x)$ is odd, $-\phi$ is solution if $\phi$ is solution; we thus assume $\phi\geq 0$ and use $\cos(\phi)=r$, $\sin(\phi)=\sqrt{1-r^2}$ to get an equation for $r$:
\begin{equation}
    \beta \left( 1-r^2 \right)^{\nu/2}f[r] = \frac{\sqrt{1-r^2}}{r}.
    \label{eq:gen_stat_r}
\end{equation}
The frozen state, corresponding to $r=1$ and $\phi=0$, is a solution for $\nu>0$.
Chiral states, corresponding to $r<1$, satisfy
\begin{equation}
    \beta f[r] = \frac{\left( 1-r^2 \right)^{(1-\nu)/2}}{r}.
    \label{eq:gen_stat_r_chiral}
\end{equation}
For $\nu<1$, there is always a solution: chiral states exist for all values of $\beta$.
For $\nu=1$, chiral states exist if $\beta>1/f[1]$.
For $\nu>1$, 0, 1 or 2 chiral states may exist, depending on the value of $\beta$.

\subsubsection{Stability analysis of the frozen state}

We perform a stability analysis of the frozen state for $\nu>0$.
We assume $r_1=r-1\ll 1$, $\phi\ll 1$ and use these expansions in eqs.~(\ref{eq:gen_overdamped_min_r}, \ref{eq:gen_overdamped_min_phi}): at the lowest order,
\begin{align}
    \dot r_1 &=-r_1-\frac{\phi^2}{2},\\
    \dot \phi &=\beta \left( r_1^2+\phi^2 \right)^{\nu/2}f \left[ \frac{\phi}{\sqrt{r_1^2+\phi^2}} \right]-\phi.
    \label{}
\end{align}
Considering positive values for $\phi$, we see that if $f(x)\geq ax$ (which is the case for the two situations considered here),
\begin{equation}
    \dot\phi\geq \left[ \beta a \left( r_1^2+\phi^2 \right)^{(\nu-1)/2}-1 \right]\phi.
    \label{eq:gen_stab}
\end{equation}
For $\nu<1$, the first term in brackets is large, meaning that the orientation dynamics is unstable.

\subsubsection{Conclusion}
We conclude that for $\nu<1$ chiral states are observed for all the values of the alignment parameter $\beta$, while for $\nu=1$ there are observed only beyond a certain value.
The situation is more complicated for $\nu>0$, as multiple chiral states exist, some of them being stable and some others being unstable.

\subsection{Self-aligning active particle in a harmonic potential with anisotropic damping}
\label{app:anisotropic_damping}

\subsubsection{General model}
The equation of motion for the position vector $\rb$ in the presence of anisotropic damping, after rescaling, is given by:
\bea
\dot{\rb} &=& \hnb - \mathbb{M} \rb~.  
\label{eom1_anisotropic}
\eea
In this context, $\mathbb{M}$ represents the mobility matrix, defined as $\mathbb{M} = \mu_{\parallel} \hnb \hnb + \mu_{\perp} (\mathbb{I}-\hnb \hnb)$. We transform mobility matrix to a single control parameter $\chi$ defined as $\mu_{\parallel}=(1+\chi)$ and $\mu_{\perp}=(1-\chi)$. For $\chi=0$, the equation of motion simplifies to the model of isotropic damping, expressed as $\dot{\rb}=\hnb-\rb$. Conversely, for $\chi=1$, it simplifies to $\dot{\rb}=[1-2(\rb\cdot\hnb)]\hnb$. Projecting on $\rb$ and $\rb^{\perp}$, we get 
\bea
\dot{r} &=& [1-(1+\chi)r\cos(\phi)]\cos(\phi) \nonumber\\
&-& (1-\chi) r \sin(\phi)\sin(\phi)\nonumber\\
r\dot{\alpha} &=& [1-(1+\chi)r\cos(\phi)]\sin(\phi)\nonumber\\
&+& (1-\chi) r \sin(\phi)\cos(\phi)
\eea
The orientation angle, evolving with self-alignment, is given by $\dot{\theta} = \beta (\hnb \times \fb)\cdot \hat{z}$, which simplifies to the following expression:
\bea
\dot{\theta} &=& -\beta (\hnb \times \rb)\cdot \hat{z}
\eea
Finally we get the $\phi = \theta-\alpha$ evolution equation
\bea
\dot{\phi} &=& \left[\frac{\beta r^2 - 1}{r}+(1+\chi)\cos(\phi)\right]\sin(\phi)\nonumber\\
&-& (1-\chi) \sin(\phi)\cos(\phi)
\eea
\subsubsection{Stationary solutions}
We obtain the stationary solutions by setting $\dot{r}=0$ and $\dot{\phi}=0$. These solutions include the frozen state, characterized by $\phi=0$ and $r=1/(1+\chi)$ and chiral states where $\phi\neq 0$.

From $\dot{r}=0$, we get $r=\cos(\phi)/[1+\chi \cos(2\phi)]$, and inserting this relation in $\dot{\phi}=0$, we get
\begin{align}
\beta \cos^2(\phi) &= (1+\chi\cos(2\phi)) [(1+\chi\cos(2\phi))\nonumber\\
&\quad -2\chi\sin(\phi)\cos^2(\phi)]
\end{align}
For isotropic damping $\chi=0$, the solutions are $\beta\cos^2(\phi)=1$ and $\beta r^2=1$. In the case of perfect anisotropic damping $\chi=1$, the solutions are $\beta=4\cos^2(\phi)(1-\sin(\phi))$ and $\beta r^2 = (1-\sqrt{1-(1/2r)^2})$.
\subsubsection{Conclusion}
The theoretical exploration of the impact of anisotropic damping on a single self-aligning particle in a harmonic potential has yet to be done.

\bibliography{sapam}

\end{document}